\documentclass[preprintnumbers,amsmath,amssymb]{revtex4}

\usepackage{natbib}

\def\br{{\bf r}}
\def\brp{{\bf {r^\prime}}}
\def\xp{{x^\prime}}

\def\ap{{\alpha^{\prime}}}
\def\aap{{a^\prime}}
\def\bp{{\beta^{\prime}}}

\def\tp{{t^\prime}}

\def\kp{{k^{\prime}}}

\def\br{{\bf r}}

\def\e{{\mbox{e}}}

\usepackage{graphicx}
\usepackage{amssymb}
\begin{document}

\title{Superoperator nonequilibrium Green's function theory of many-body systems; Applications to charge transfer
and transport in open junctions}

\author{U. Harbola and S. Mukamel}
\affiliation{\em Department of Chemistry, University of California, Irvine, California 92697}

\begin{abstract}
Nonequilibrium Green's functions provide a powerful tool for
computing the dynamical response and particle exchange statistics of coupled
quantum systems. We formulate the theory in terms of the
density matrix in Liouville space and introduce superoperator algebra that
greatly simplifies the derivation and the physical interpretation of all quantities.
Expressions for various
observables are derived directly in real time in terms of superoperator nonequilibrium Green's functions
(SNGF), rather than the artificial time-loop
required in Schwinger's Hilbert-space formulation.
Applications for computing interaction energies, charge densities,
average currents, current induced fluorescence, electroluminescence
and current fluctuation (electron counting) statistics are discussed.\\

\end{abstract}

\maketitle

\section{Introduction}
Nonequilibrium Green's function theory (NEGFT)
\cite{schwinger,keldysh,langreth,rammer-book07,kadanoff-baym,alexandre,mahan-book}
is widely used for computing electron transport and
optical properties of open many-body
systems, such as semiconductors\cite{haug-jauho}, metals \cite{rammer-smith},
molecular wires and scanning tunneling microscopy (STM) junctions\cite{datta-book}.
It has also been applied to X-ray photoemission spectroscopy
\cite{hirokoPRB72,caroliPRB8,fujikawaCPL368}. General
fluctuation-dissipation relations may be derived for
nonlinear response and fluctuations using this formalism \cite{chou,wang}.
In NEGFT, the diagrammatic perturbative expansions are carried
on a closed time-loop which includes two branches with forward and
backward time propagation, respectively. The closed-time-path was first
introduced by Schwinger \cite{schwinger} and further developed by
Keldysh and others \cite{keldysh,kadanoff-baym,mills}.
The physical time $t$ is replaced by a loop-time $\tau $
which runs clockwise on a contour (Fig.1a) such that when $\tau $
goes from $\tau _{i}$ to $\tau _{f}$, the physical time $t$ runs first forward
and then backward. This technique allows to
treat nonequilibrium systems using quantum field theory tools originally developed for
equilibrium systems.

The NEGFT has been remarkably successful in describing a broad range of non-equilibrium
systems and  phenomena. Frequency domain non-linear
susceptibilities can be readily interpreted by an evolution on the
loop \cite{mukamelPRA}.
However loop approach does not offer an obvious  physical real-time intuition for the
various quantities and approximations. This is only possible at the end of the calculation,
when one transforms $\tau $ to real time.
A highly desirable alternative, real-time, formalism has been
developed using two time
ordering prescriptions known as the "single time" and the "physical"
representations \cite{hao,chou,wai-yong,wang}. For reasons that will become
clear later, we shall denote these the PTBK (Physical-time bra-ket) and the
PTSA (Physical-time symmetric-antisymmetric) representations, respectively.
We shall refer to Schwinger's formulation as the closed time path loop (CTPL).
In Ref. \cite{chou}
NEGFT for bosons was formulated in the PTSA representation by
considering different evolutions in the forward and backward branches of
the time-loop. A generating functional for the
Keldysh Green's functions was constructed by introducing different artificial fields which
couple to operators in the forward and the backward branches of the
loop. This is similar to the Liouville space \cite{shaul-book}
formulation of NEGF presented in Ref.
\cite{up-shaul-JCP} where  Hedin's equations \cite{hedin,onida-RMP,gunnarsson-Review}
were generalized to an open system. The PTSA representation was
finally obtained in Ref. \cite{chou} in two steps;
first formulating the theory in the PTBK and subsequently
making a matrix transformation on the generating functional to obtain the PTSA representation.

In this article we show how by formulating the many-body problem
using superoperators in Liouville space, the PTSA representation can be
used from the outset without introducing any artificial fields. The matrix
transformation between the two representations
is performed on the superoperators themselves rather
than on some expectation values (Green's functions or generating
functions). The time evolution of superoperators in the PTSA representation
can then be calculated directly from the microscopic equations of motion for
superoperatrors, totally avoiding the intermediate PTBK representation.
The observables are computed  directly
in terms of the retarded, advanced and correlation superoperator nonequilibrium Green's functions (SNGF).
For completeness, we introduce bosonic superoperators and outline the superoperator approach for bosons.

Applications are made to inelastic resonances in STM currents
\cite{up-shaul-prb1} and current induced fluorescence in STM
junctions \cite{up-jeremy-prb2}. A key ingredient of
the SNGF approach is a simple time-ordering prescription of superoperators
that provides an intuitive and powerful bookkeeping device for all interactions, and
maintains a physical picture based on the density-matrix. The physical significance of the various Green's becomes obvious.

In contrast to Hilbert space CTPL which targets the wave function, the
Liouville space PTBK and PTSA formulations are based on the density matrix.
The backward evolution in Hilbert
space is replaced by the simultaneous evolution of the bra and ket
of the density matrix, which can have a different evolution (but always
forward in time!) in Liouville space (Fig. 1b).
Our notation appears naturally when using
the density matrix. The "single time", PTBK, works with
superoperators which act on the bra and the ket.
In the "physical", PTSA, representation which most directly resembles some observables,
we work with linear combinations (sums and differences) of the PTBK superoperators.

The SNGF has been first developed and applied to {\em closed}
interacting many-body systems;
nonlinear optical response of excitons in molecular aggregates \cite{chernyak} and
non-equilibrium Van der Waals forces \cite{cohen-mukamel-1,cohenPRL}.
Properties of the
interacting system were determined by the density response as well
as the correlated density fluctuations
\cite{up-shaulPRA,cohen-mukamel-1} of the individual
sub-systems. The same approach has been used to resolve the causality
paradox \cite{shaulPRA,van-leeuwen} of density-functional
theory\cite{kohn-sham-PhyRev1965,kohn-PhyRev1964}. Here we extend this
method to open many-body systems with overlapping charge
densities so that electron transfer is possible and the number of
particles in each subsystem can fluctuate. The SNGF appear naturally as a consequence of the time
ordering of the ket and the bra evolution of the density matrix
which gives rise to {\em Liouville space pathways} (LSP)\cite{shaul-book}. Each
SNGF represents a particular combination of LSPs. This
establishes a connection between many-body nonequilibrium theory and
the standard formulation of nonlinear optical response
\cite{shaul-book} and paves the way for computing nonlinear optical
properties of quantum junctions.

Many types of charge-transfer processes are possible in open
systems. The simplest is between two bound states. These are the
fundamental processes in many chemical reactions,
mixed-valence complexes \cite{voorhis-1} and
biological processes, such as photosynthesis
and cell respiration \cite{marcus,jortner}. A second type of electron transfer
occurs between two coupled semi-infinite many-electron sub-systems $A$ and $B$
held at different chemical potentials.
Direct tunneling takes place between two quasi-free states subjected to a thermodynamic
driving force stemming from the differences in chemical potentials.
Nonequilibrium single-electron transfer statistics between two leads has been
studied extensively over the past two decades since it reveals the fundamental
quantum effects and direct applications to nanodevices\cite{hanbury-brown,hanbury-brown-boson}.
Single electron counting has many
similarities to the more mature field of single photon counting \cite{glauber-book,mukamel-counting}.
The steady state current between $A$ and $B$ can be
computed using the Landauer-Buttiker (LB) scattering matrix formalism \cite{lb,datta-book}
\begin{equation}
\label{nn-44}
I=\frac{e}{\pi}\int_{0}^{\infty
}dE[f_A(E)-f_B(E)]\sum_{ab}|S_{ab}(E)|^{2}
\end{equation}
where $f_X(E)=[1+e^{\beta(E-\mu_X)}]^{-1}$ is the Fermi function for
lead, $X=A,B$ with chemical potential $\mu_X$, and $S_{ab}$ is
a scattering matrix element for electron transmission from $a$
-th mode on $A$ to the $b$-th mode on lead $B$. We use $\hbar=1$.

Bardeen's perturbative approach \cite{bardeen} has been very successful for imaging metal
surfaces in scanning tunneling microscopy (STM).
Tursoff and Hamman \cite{TH} had used it for computing the current in STM configurations.
Using a spherical tip geometry and a generic wavefunction for the metal
surface which decays exponentially in the tunneling region, they arrived at
a simple formula for tunneling current
\begin{eqnarray}  \label{TH-form}
\frac{dI}{dV} \propto \sigma(r_0,E_F)
\end{eqnarray}
where $\sigma$ is the density of states (DOS) of the metal surface
at the tip position $r_0$ and the equilibrium energy $E_F$.

\begin{figure}[h]
\centering
\rotatebox{0}{\scalebox{.8}{\includegraphics{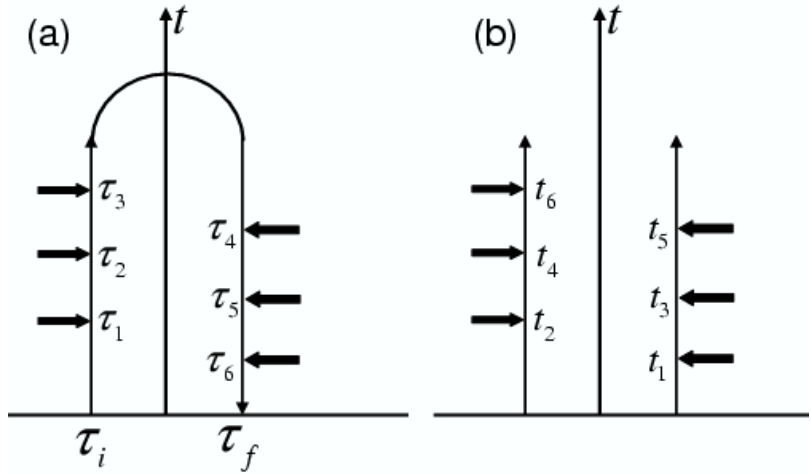}}}\newline
\caption{\it Time contour for Eq. (\protect\ref{expec-2}). Physical time
($t$) runs from bottom to top.
Arrows at different times indicate interactions
of the bra or the ket of the density matrix with an external perturbation.
a) Evolution in Hilbert space involves first interacting with the ket and then
with the bra. Loop time variables ($\tau$) are ordered clockwise,
$\tau_1<\tau_2<\tau_3<\tau_4<\tau_5<\tau_6$.
Interactions are time-ordered in physical time within each branch but not between
branches (partial time ordering\cite{marx}).
b) In Liouville space, both bra and ket evolve forward. All interactions are
ordered in physical time ($t_1<t_2<t_3<t_4<t_5<t_6$).
}
\label{fig3}
\end{figure}

Electron transport through a quantum system,
$e.$$g.$ a single molecule or a chain of atoms, connected to two
macroscopic leads ($A$ and $B$) constitutes a third type of charge transfer.
In this "molecular-wire" configuration, the electron moves sequentially
from lead $A$ to the system
and then to lead $B$$^{1}$. \footnotetext[1]{ Note
that there is a finite probability for the electron to return to
the first lead without tunneling across the junction. Such processes are relevant for the current statistics applications at low bias discussed in Section \ref{statistics}.} In contrast to the tunneling case, where each of the
two leads is held at equilibrium with its own chemical potential,
this case is more complex since it involve a non-equilibrium state of the embedded quantum
system.

Electron coupling with molecular vibrations may result in inelastic
scattering. These processes contain signatures of
the bonding environment and the nonequilibrium phonon states of
molecules in STM junctions \cite{w-ho}. Such processes are not
captured by the simple LB scattering picture.
The early NEGF formulation of currents through molecular
junctions developed by Caroli $et$ $al$ \cite{caroli} took into
account inelastic processes and the nonequilibrium state of
the quantum system. This approach
has since been broadly applied to study the conductance properties of molecules in
STM \cite{up-shaul-prb1,stm,LorentePRL2000,blanco} and molecular wire junctions
\cite{molecular-current,Galperin}, and to the optical properties of
semiconductors \cite{haug-jauho}.
Recently Cheng $et$ $al$ \cite{voorhis-2} have used the time-dependent density-functional
theory to numerically compute the conductance of a polyacetylene molecular wire and demonstrated that the NEGF theory works well.

Current-carrying molecule may show a negative conductance. The change in resonance
condition between two conducting orbitals from different parts of a
molecule can influence the conductance significantly, which, under
certain conditions, may induce negative conductance \cite{neg-cond,diventra-1}.
Inelastic effects
can also lead to negative conductance. This was demonstrated by
Galperin $et$ $al$ \cite{Galperin} using the self-consistent
solution for the NEGF in the presence of electron-phonon interactions.
Recently, Maddox $et$ $al$ \cite{maddox} have used a many-body
expansion of the SNGF to explain the hysteresis switching behavior
observed \cite{WuScience} in a single magnesium porphine molecule
absorbed on NiAl surface in an STM junction. In order to analyze
inelastic effects in STM imaging of single molecules, Lorente and
Persson \cite{LorentePRL2000} combined the Tersoff-Hamann theory with
the formulation of Caroli $et$ $al$\cite{caroli} to compute the fractional change in
the DOS, Eq. (\ref{TH-form}), perturbatively in the electron-phonon
coupling using NEGF.

The article is organized as follows. The various prescriptions for bookkeeping
for time-ordering are first introduced without alluding to many-body theory
in sections \ref{sk-loop} and \ref{superoperator}: The Hilbert-space CTPL
technique is presented in Sec. \ref{sk-loop} and in Sec. \ref{superoperator}
we consider the superoperators and their algebra in Liouville space and introduce the PTBK
and PTSA prescriptions. In Sec. \ref{closed-systems} we treat closed interacting systems
using the superoperator approach. This illustrates how superoperator time-ordering works.
Application is made to Van der Waals' forces. The remainder of the review presents applications to {\em open} many-body systems.
In Sec. \ref{negf-fermion}, we introduce Fermi superoperators.
In Sec. \ref{gf-nn} we define the many-body SNGF and connect them to
retarded, advanced and correlation Green's functions.
In Sec. \ref{observables}, we recast various observables such as the interaction energy,
the charge density profile, current induced fluorescence and
electroluminescence of interacting sub-systems in terms of these
Green's functions. In Sec. \ref{sec-dyson}, a closed matrix Dyson equation is derived for
the SNGF which allows to include electron-electron or
electron-phonon interactions perturbatively
through a self-energy matrix. This is reminiscent of the
Martin-Siggia-Rose\cite{msr} formulation in classical mechanics.
Only three Green's functions appear in the PTSA formulation (the fourth vanishes
identically). The calculation is simplified considerably since the matrix Dyson equation decouples from the
outset and the retarded, advanced and correlation Green's functions
can be calculated separately.
In the CTPL approach, in contrast, we must first calculate
four Keldysh Green's functions and then perform a linear transformation
to obtain the PTSA in terms of the
retarded, advanced and correlation Green's functions. This two-step approach is not
required in the Liouville-space formulation as the microscopic equations of motion can be
constructed directly for the PTSA operators. Current-fluctuation statistics in tunneling
junctions provides a wealth of information. Extending it to the single
electron counting regime (as is commonly done for photon counting \cite{glauber-book,mukamel-counting})
is an important recent development.
This is connected to the SNGF in Sec \ref{statistics}. Superoperator algebra for
bosons is introduced in Appendix \ref{boson}\cite{jansen}. Other appendices give technical details.
We finally conclude in Sec. \ref{conclusion}.

\section{Schwinger's Closed Time Path Loop  (CTPL) Formalism}\label{sk-loop}

We consider a system described by the Hamiltonian,
\begin{equation}
\label{hamil-1}
\hat{H}_{T}(t)=\hat{H}+\hat{V}(t),
\end{equation}
where $\hat{V}(t)$ is an external perturbation that drives the system out of
equilibrium, and
\begin{eqnarray}
\hat{H}=\hat{H}_{0}+\hat{H}^{\prime }
\end{eqnarray}
is a sum of a zero order
part $(\hat{H}_{0})$ and an interaction part $H^{\prime }$ which will be
treated perturbatively. For $t<t_{0}$, $V(t)=0$,
and the system is described by the canonical density matrix,
\begin{equation}
\label{rho-1}
\hat{\rho}(t_0)=\frac{\mbox{e}^{-\beta \hat{H}}}{Tr\{\mbox{e}^{-\beta \hat{H}}\}}.
\end{equation}

We are interested in computing the expectation value of an operator $\hat{\mathcal{O}}$
\begin{eqnarray}
\label{aaa}
S=Tr[\hat{\mathcal{O}}\hat{\rho}(t)],~~~~  t>t_{0}.
\end{eqnarray}
We work in the Schrodinger picture where
\begin{eqnarray}
\label{eom-1}
\hat{\rho}(t) = \hat{U}(t,t_0)\hat{\rho}(t_0)\hat{U}^\dag(t,t_0)
\end{eqnarray}
with the time evolution operator
\begin{eqnarray}
\label{aaa-1}
\hat{U}(t,t_0)&=& \hat{T}\mbox{exp}\left[-i\int_{t_0}^td\tau \hat{H}_T(\tau)\right],
\end{eqnarray}
and its hermitian conjugate
\begin{eqnarray}
\label{aaa-2}
\hat{U}^\dag(t,t_0)&=& \hat{T}^\dag\mbox{exp}\left[i\int_{t_0}^td\tau \hat{H}_T(\tau)\right].
\end{eqnarray}
Here $\hat{T}(\hat{T}
^{\dag })$ are the forward (backward) time ordering operators: when acting
on a product of operators they rearrange them in increasing (decreasing) order of time from right to left.

We next switch from the
Schrodinger to the interaction picture where
the time dependence of an operator  $\hat{\mathcal{O}}(t)$
is given by
\begin{eqnarray}
\label{eom-3}
 \hat{\mathcal{O}}_I(t) = \hat{U}^{\dag}_H(t,t_{0})\hat{\mathcal{O}}(t_0)\hat{U}_H(t,t_{0})
\end{eqnarray}
here
\begin{eqnarray}
\hat{U}_H(t,t_0)&=& \theta(t-t_0)\mbox{exp}\{-i\hat{H}(t-t_0)\}
\end{eqnarray}
represents the evolution with respect to $\hat{H}$ rather than $\hat{H}_{T}$.
The density-matrix evolution in the interaction picture is given by,
\begin{equation}
\label{inter-time}
\hat{\rho}_I(t)= \hat{U}_I(t,t_{0})\hat{\rho}(t_0)\hat{U}_I^\dag(t,t_{0}),
\end{equation}
where
\begin{eqnarray}
\hat{U}_I(t,t_{0}) &=&\hat{T}\mbox{exp}\left\{ -i\int_{t_{0}}^{t}d\tau \hat{V}_I
(\tau )\right\}, \nonumber  \label{time-evo-1} \\
\hat{U}^{\dag }_I(t,t_{0}) &=&\hat{T}^{\dag }\mbox{exp}\left\{ i\int_{t_{0}}^{t}\hat{
V}_{I}(\tau )\right\}
\end{eqnarray}
are the interaction picture propagators.
Equation (\ref{inter-time}) can be recast as
\begin{equation}
\hat{\rho}_I(t)=\hat{T}_{C}\mbox{exp}\left\{ -i\int_{C}d\tau \hat{V}
_{I}(\tau )\right\} \hat{\rho}(t_0)
\end{equation}
where $\hat{T}_{C}$ denotes time-ordering on the contour $C$ shown in Fig.
1a. Our observable, Eq. (\ref{aaa}), finally becomes
\begin{equation}
\label{expec-2a}
S=\left\langle \hat{T}_{C}\mbox{exp}\left\{ -i\int_{C}d\tau \hat{V}_{I}(\tau
)\right\} \hat{\mathcal{O}}_{I}(t)\right\rangle .
\end{equation}
Here $\langle\cdots\rangle=\mbox{Tr}[\cdots\hat{\rho}(t_0)]$ is the average
with respect to the density matrix Eq. (\ref{rho-1}) which includes the interactions $\hat{H}^\prime$.
For practical applications, one has to convert this to the expectation value with respect
to $H_{0}$. The contour in Fig. 1a is then modified to account for
the initial correlations \cite{haug-jauho,wagner,rammer-smith} which
are important only for the transient properties. In most
physical problems, like non-equilibrium steady state, this modification is not
important. Note that the time dependence on $H$ appears at three places in Eq. (\ref{expec-2a})
[in $\hat{V}_I(t), \hat{\mathcal{O}}_I(t)$ and $\rho(t_0)$].
Using a second interaction picture, we can switch the time dependence
from $H$ to $H_0$.
The density matrix $\rho(t_0)$ is obtained by adiabatic (Gell-Mann law) switching of the
interaction starting at $t\to -\infty$, where the system is described by $H_0$ alone.
\begin{eqnarray}
\label{adiabatic}
 \rho(t_0)=\bar{U}_I(t_0,-\infty)\rho_0\bar{U}^\dag_I(t_0,-\infty)
\end{eqnarray}
$\rho_0$ is the density matrix of the non-interacting system
(Eq. (\ref{rho-1}) with $\hat{H}$ replaced by $\hat{H}_0$) and
\begin{eqnarray}
\bar{U}_I(t_0,-\infty)&=& \hat{T}\mbox{exp}
\left\{-i\int_{-\infty}^{t_0}d\tau \hat{H}^\prime_{H_0}(\tau)\right\}\\
 \hat{H}^\prime_{H_0}(\tau)&=& \hat{U}^\dag_{H_0}(\tau,0)\hat{H}^\prime \hat{U}_{H_0}(\tau,0)
\end{eqnarray}
with $\hat{U}_{H_0}(t,0)=\theta(t)\mbox{e}^{-i\hat{H}_0t}$ is the time evolution with respect to $H_0$.
Equation (\ref{eom-3}) can be similarly expressed as,
\begin{eqnarray}
\label{eom-4}
 \hat{\mathcal{O}}_I(t) = \bar{U}^{\dag}_I(t,t_0)\hat{\mathcal{O}}_{H_0}(t)\bar{U}_I(t,t_{0}).
\end{eqnarray}

Substituting Eqs. (\ref{adiabatic}) and (\ref{eom-4}) in (\ref{expec-2a}) and combining
the exponentials (this can be done thanks to the time ordering operators which allows us to
move operators within a product of operators), we finally obtain
\begin{equation}
\label{expec-2}
S=\left\langle \hat{T}_{C}\mbox{exp}\left\{ -i\int_{C}d\tau \hat{\cal W}_{H_0}(\tau
)\right\} \hat{\mathcal{O}}_{H_0}(t)\right\rangle_0 .
\end{equation}
where $\tau$ varies on a contour $C$ which starts at $t=-\infty$ and
$\hat{\cal W}_{H_0}(t)=\hat{H^\prime}_{H_0}(t)+\hat{V}_{H_0}(t)$.
All time dependence is now given by $\hat{H}_0$ and $\langle\cdots\rangle_0$ is the trace with respect
to $\rho_0$.
This formally looks like  equilibrium
(zero or finite temperature) theory. The only difference is that
the real time integrals are replaced by contour integrals.
Thus in the CTPL approach, the non-equilibrium theory is mapped onto the
equilibrium one, where standard Feynman diagram techniques and Wick's theorem
can be applied.

\section{Liouville-space superoperators and their algebra}\label{superoperator}

The $N\times N$ density matrix
in Hilbert space is written in Liouville space \cite{fano,zwanzig-book,revven} as a vector of length $N^{2}$.
With any Hilbert space operator
$\hat{A}$ we associate two Liouville space
superoperators labeled "left", $\breve{A}_{L}$, and "right", $\breve{A}_{R}$. These are defined by their actions on any other operator $X$ \cite{chernyak,shaul-PRE}
\begin{equation}
\label{liouv-operator}
\breve{A}_{L}X\equiv \hat{A}X,~~~\breve{A}_{R}X\equiv X\hat{A}.
\end{equation}

We also introduce the unitary transformation
\begin{equation}
\label{trans}
\breve{A}_{+}=\frac{1}{\sqrt{2}}(\breve{A}_{L}+\breve{A}_{R}),~~~\breve{A}
_{-}=\frac{1}{\sqrt{2}}(\breve{A}_{L}-\breve{A}_{R}).
\end{equation}
The inverse transformation can be
obtained simply by interchanging $+$ and $-$ with $L$ and $R$, respectively.
In the following expressions we denote
superoperator indices by Greek letters($\alpha ,\beta, \kappa,\eta$). These can be
either $+,-$ (the PTSA representation) or $L,R$ (the PTBK representation).
In the $+,-$ representation a single act of $A_{-}$ in Liouville space
represents the commutation with $A$ in the Hilbert space. Thus the
nested commutators that appear in perturbative expansions in Hilbert
space transform to a compact notation which are more easy to interpret in
terms of the double sided Feynman diagrams\cite{shaul-book}. Similarly, a
single action of $A_{+}$ in Liouville space corresponds to an
anticommutator in Hilbert space.
\begin{eqnarray}
\breve{A}_{-}\hat{X} &\equiv &\frac{1}{\sqrt{2}}(\hat{A}\hat{X}-\hat{X}\hat{A})
\label{transform} \\
\breve{A}_{+}\hat{X} &\equiv &\frac{1}{\sqrt{2}}(\hat{A}\hat{X}+\hat{X}\hat{A}).
\end{eqnarray}

Note that in the "single time" nomenclature \cite{chou,wang}
one often denotes the left branch by $+$ (positive time for the loop) and the
right branch by $-$ (negative). Here we denote these by $L$ and $R$. The "plus" ($+$) and "minus" ($-$) are reserved for the sum and difference combinations in the PTSA representation.
We believe that this notation does most justice to the significance of the two pictures within the density matrix framework.

With any product of operators in Hilbert space, we can associate
corresponding superoperators in Liouville space.
\begin{eqnarray}
\label{nnn-1}
(\hat{A}_{i}\hat{A}_{j}\cdots \hat{A}_{k})_{L} &=&\breve{A}_{iL}\breve{A}_{jL}\cdots \breve{A}
_{kL}  \nonumber  \label{identity} \\
(\hat{A}_{i}\hat{A}_{j}\cdots \hat{A}_{k})_{R} &=&\breve{A}_{kR}\cdots \breve{A}_{jR}\breve{A}
_{iR}.
\end{eqnarray}
This follows directly from the basic definition, Eq. (\ref{liouv-operator}).

The following rules which immediately follow from Eqs. (\ref{trans}) and
(\ref{nnn-1}) allow to convert
products of operators directly to products of $+/-$ superoperators,
\begin{eqnarray}
(\hat{A}_{i}\hat{A}_{j})_{-} &=&\frac{1}{2\sqrt{2}}\left[ [\breve{A}_{i+},\breve{A}%
_{j+}]+[\breve{A}_{i-},\breve{A}_{j-}]\right.   \nonumber  \\
&+&\left. \{\breve{A}_{i+},\breve{A}_{j-}\}+\{\breve{A}_{i-},\breve{A}_{j+}\}%
\right] \label{pm-1} \\
(\hat{A}_{i}\hat{A}_{j})_{+} &=&\frac{1}{2\sqrt{2}}\left[ \{\breve{A}_{i+},\breve{A}%
_{j+}\}+\{\breve{A}_{i-},\breve{A}_{j-}\}\right.   \nonumber \\
&+&\left. [\breve{A}_{i+},\breve{A}_{j-}]+[\breve{A}_{i-},\breve{A}_{j+}]
\right]. \label{pm}
\end{eqnarray}
Equations (\ref{pm-1}) and (\ref{pm}) may be used to recast functions of Hilbert space
operators, such as the Hamiltonian, in terms of superoperators in Liouville space.

The time-ordering operator $\breve{T}$ is a key tool in the Liouville space formalism;
when acting on a product of time-dependent superoperators, it
rearranges them in increasing order of time from right to left.
\begin{eqnarray}
\label{two-time-ordering}
\breve{T}\breve{A}_{i\alpha }(t)\breve{A}_{j\beta }({t^{\prime }})=\theta(t^\prime-t)
\breve{A}_{j\beta }({t^{\prime }})\breve{A}_{i\alpha }(t)+
\theta(t-t^\prime)\breve{A}_{i\alpha }(t)\breve{A}_{j\beta }({t^{\prime }})
\end{eqnarray}
Note that, unlike Hilbert space where we
need two time-ordering operators to describe the evolution in
opposite forward ($\hat{T}$) and backward ($\hat{T}^\dag$) directions (Eqs. (\ref{aaa-1}) and (\ref{aaa-2})), the Liouville space operator $\breve{T}$ always acts
to its right and only forward propagation is required. This facilitates the physical
real-time interpretation of all algebraic expressions obtained in
perturbative expansions.

Using Eqs. (\ref{liouv-operator}) and (\ref{trans}), it is straightforward
to show that
\begin{eqnarray}
\label{liu-hil}
\langle \breve{A}_{L}\rangle  &=&\langle \breve{A}_{R}\rangle \equiv \langle
\hat{A}\rangle   \nonumber  \\
\langle \breve{A}_{+}\rangle  &\equiv &\sqrt{2}\langle \hat{A}\rangle
,~~~~\langle \breve{A}_{-}\rangle =0.
\end{eqnarray}
where $\langle \cdot \rangle $ represents a trace which includes the
density matrix, $\langle A\rangle$=Tr$\{A\rho\}$.
$\langle \breve{A}_-\rangle$ vanishes since it is the tract of a commutator.
An important consequence of the last equality in
Eq. (\ref{liu-hil}) is that the average of a product of "plus"($+$) and "minus"
($-$) operators with left most "minus" operator vanishes,
i.e. $\langle \breve{A}_{1-}\cdots\breve{A}_{n\alpha}\rangle=0$ and therefore
the average of a product of "all-minus" operators vanishes identically
$\langle \breve{A}_{1-}\breve{A}_{2-}\cdots \breve{A}_{n-}\rangle=0$.
The following time-ordered average of two superoperators is therefore causal
\begin{eqnarray}
\label{causality}
\langle \breve{T}\breve{A}_{1+}(t_1)\breve{A}_{2-}(t_2)\rangle =\left\{
\begin{array}{cc}
\langle \breve{A}_{1+}(t)\breve{A}_{2-}(t_2)\rangle,& ~~~~ t_1>t_2\\
\langle \breve{A}_{2-}(t_2) \breve{A}_{1+}(t)\rangle =0&~~~~~t_2>t_1.
\end{array}
\right.
\end{eqnarray}

For fermion operators $\breve{A}_{R}$ is defined differently than
in Eq. (\ref{liouv-operator}) in order to maintain simple
superoperator commutation rules. Consequently, some of the above results, in particular Eqs. (\ref{liu-hil}), no longer hold. Fermion superoperators will be discussed in Sec. \ref{negf-fermion}.

\subsection{The interaction-picture}

We are interested in computing the expectation value, Eq. (\ref{aaa}), for the system described by the Hamiltonian, Eq. (\ref{hamil-1}).
The superoperators corresponding to $\hat{H}$, $\hat{H}_{0}$,$\hat{V}$
and $\hat{{\cal W}}$ will be denoted as $\breve{H}$,
$\breve{H}_{0}$, $\breve{V}$ and $\breve{\cal W}$, respectively.

The time-evolution of the density matrix in the Schrodinger picture is
given by the Liouville equation
\begin{equation}
\frac{\partial }{\partial t}\hat{\rho}(t)=-i[\hat{H}_T(t),\hat{\rho}(t)]
\end{equation}
which in superoperator notation reads,
\begin{equation}
\label{liouville-time-evolution}
\frac{\partial}{\partial t}\hat{\rho }(t) =-i\sqrt{2}\breve{H}_{T-}\hat
{\rho}(t).
\end{equation}
The formal solution of Eq. (\ref{liouville-time-evolution}) is
\begin{equation}
\hat{\rho}(t) ={\breve{U}}(t,t_{0})\hat{\rho }(t_0)
\end{equation}
where
\begin{equation}
\label{evolution-full}
{\breve{U}}(t,t_{0})=\breve{T}\mbox{exp}\left\{-i\sqrt{2}\int_{t_0}^t d\tau
{\breve{H}_{T-}}(\tau)\right\}\hat{\rho }(t_0)
\end{equation}
is the time evolution operator in Liouville space.

By substituting Eq. (\ref{hamil-1}) in (\ref{evolution-full}), we can switch
to the interaction picture
\begin{equation}
\label{parti}
{\breve{U}}(t,t_{0})={\breve{U}}_{0}(t,t_{0}){\breve{U}}_I(t,t_{0}),
\end{equation}
where
\begin{equation}
{\breve{U}}_{0}(t,t_{0})=\theta (t-t_{0})\mbox{e}^{-i\sqrt{2}{\breve{H}_{0-}}(t-t_{0})},
\end{equation}
represents the free evolution, and
\begin{equation}
{\breve{U}}_I(t,t_{0})=\breve{T}\mbox{exp}\left\{ -i\sqrt{2}\int_{t_{0}}^{t}d\tau {
\breve{\cal W}^I_{-}}(\tau )\right\}.
\end{equation}
In the interaction picture, the time dependence of any superoperator is given by,
\begin{equation}
\breve{\cal W}^I_{\alpha }(t)={\breve{U}}_{0}^{\dag}(t,t_0)\breve{\cal W}_{\alpha}(t_0){\breve{U}}_{0}(t,t_0).
\end{equation}
where $\alpha =L,R$ or $+,-$. Interaction picture superoperators will be denoted by a superscript $I$, $\breve{A}^I_\alpha$.
Using this notation, our observable, Eq. (\ref{aaa}), is then given by
\begin{eqnarray}
\label{ex-1}
S=\mbox{Tr}[\breve{\cal O}_I(t)\hat{\rho}_I(t)]
\end{eqnarray}
where $\hat{\rho}_I(t)=\breve{U}_I(t,t_0)\hat{\rho}(t_0)$ is the density
matrix in the interaction picture.
$\hat{\rho}_I$ can be generated from the density matrix, $\rho_0$, of the
non-interacting system by switching the interaction $\hat{\cal W}$ adiabatically starting at $t_0=-\infty$.
\begin{equation}
\label{adiabatic-1}
\hat{\rho}_I(t)=\breve{U}_I(t,-\infty )\hat{\rho}_0
\end{equation}
Combining Eqs. (\ref{adiabatic-1}) and (\ref{ex-1}) gives
\begin{eqnarray}
S = \left\langle\breve{T}\mbox{exp}\left\{-i\sqrt{2}\int_{-\infty}^t \breve{\cal
W}^I_{-}(\tau)d\tau\right\} \breve{{\cal O}_I}(t)\right\rangle_0
\end{eqnarray}
In contrast to Eq. (\ref{expec-2}), we note that we only
need a single time-ordering operator which is always defined in
real-time. By expanding the exponential, $S$ can
be computed perturbatively in the interaction $V$.

\section{Superoperator formalism for closed interacting systems: Van der Waals' forces}\label{closed-systems}

Here we demonstrate how the superoperator approach is used by applying it to intermolecular forces \cite{cohen-mukamel-1,cohenPRL,up-shaulPRA}. This "single-body" example sets the stage for the many-body applications presented in the remainder of this review.

Consider two interacting systems of charged particles, $A$ and $B$, kept at sufficiently large distance such that their charges do not
overlap and their interaction is purely Coulombic.
The total Hamiltonian is
\begin{eqnarray}
\label{close-1}
 \hat{H}=\hat{H}_A+\hat{H}_B+\hat{H}_{AB}
\end{eqnarray}
where $\hat{H}_A$ $\hat{H}_B$ represent the isolated systems and the coupling is
\begin{eqnarray}
\label{close-2}
\hat{H}_{AB} = \int d\br\int d\brp K(|\br-\brp|)\hat{Q}_A(\br)\hat{Q}_B(\brp)
\end{eqnarray}
where $\hat{Q}_X(\br)$ is the charge-density operator for system $X$ at point $\br$
and $~~~~$ $K(|\br-\brp|)=1/|\br-\brp|$ is the coulomb potential.

The interaction energy is defined as the change in the total energy of the system due to
interaction $\hat{H}_{AB}$, of the system.
\begin{eqnarray}
\label{close-3}
\Delta E &=& \langle \hat{H}_{AB} \rangle_{AB}
- \langle \hat{H}_{A} \rangle_{A}-\langle \hat{H}_{B} \rangle_{B}\nonumber\\
\end{eqnarray}
where $\langle\cdots\rangle_{AB}$ is the trace with respect to the ground state density-matrix of the interacting system and $\langle\cdots\rangle_{A}$ is the trace
with respect to noninteracting density-matrix of $A$, and similarly for $B$.

We introduce a switching parameter $0<\eta<1$ and define the Hamiltonian
\begin{eqnarray}
\label{eta-hamil}
\hat{H}_\eta = \hat{H}_A+\hat{H}_B+\eta\hat{H}_{AB}.
\end{eqnarray}
Using the Hellman-Feynman theorem,
Eq. (\ref{close-3}) gives
\begin{eqnarray}
\label{close-4}
\Delta E &=& \int_0^1 d\eta \langle \hat{H}_{AB}\rangle_\eta
\equiv \frac{1}{\sqrt{2}}\int_0^1 d\eta \langle \breve{H}_{AB+}\rangle_\eta
\end{eqnarray}
where $\langle\cdots\rangle_\eta$ denotes a trace with respect to the density-matrix
corresponding to the Hamiltonian $H_\eta$.
The integral in Eq. (\ref{close-4}) is due
to the adiabatic switching of parameter $\eta$ from zero (noninteracting systems) to
one (when $\hat{H}_{AB\eta}=\hat{H}_{AB}$). In going from first equality to second
we have used  Eq. (\ref{liu-hil}).

The superoperators, $\breve{H}_{AB\alpha}, \alpha=+,-$ corresponding to the coupling, $\hat{H}_{AB}$, are obtained using Eqs. (\ref{pm-1}) and (\ref{pm}) in Eq. (\ref{close-2}). Since $\hat{Q}_A$ and $\hat{Q}_B$ commute, we get
\begin{eqnarray}
\label{close-5}
\breve{H}_{AB-} &=& \int d\br\int d\brp K(|\br-\brp|)(\hat{Q}_A(\br)\hat{Q}_B(\brp))_-\nonumber\\
&=& \frac{1}{\sqrt{2}}\int d\br\int d\brp K(|\br-\brp|)[\breve{Q}_{A+}(\br)\breve{Q}_{B-}(\brp)
+\breve{Q}_{A-}(\br)\breve{Q}_{B+}(\brp)].
\end{eqnarray}
Similarly,
\begin{eqnarray}
\label{close-6}
\breve{H}_{AB+} &=& \int d\br\int d\brp K(|\br-\brp|)(\hat{Q}_A(\br)\hat{Q}_B(\brp))_+\nonumber\\
&=& \frac{1}{\sqrt{2}}\int d\br\int d\brp K(|\br-\brp|) [\breve{Q}_{A+}(\br)\breve{Q}_{B+}(\brp)
+\breve{Q}_{A-}(\br)\breve{Q}_{B-}(\brp)].
\end{eqnarray}

Substituting Eq. (\ref{close-6}) in (\ref{close-3}) and using the fact that
$\langle \breve{Q}_{A-}(\br)\breve{Q}_{B-}(\brp)\rangle=0$, we obtain
\begin{eqnarray}
\label{close-7}
\Delta E = \frac{1}{2} \int_0^1 d\eta \int d\br\int d\brp
K(|\br-\brp|)\langle \breve{Q}_{A+}(\br)\breve{Q}_{B+}(\brp)\rangle_\eta .
\end{eqnarray}

We next define the charge-density fluctuation operator corresponding in
system $X$, $\delta \hat{Q}_X(\br,t)=\hat{Q}_X(\br,t)-\bar{Q}_X(\br)$, where $\bar{Q}(\br)$ is the average charge-density at point $\br$. The total interaction energy in
Eq. (\ref{close-7}) can then be factorized into three parts,
$\Delta E = \Delta E_0+\Delta E_1 +\Delta E_2$, where
\begin{eqnarray}
\label{close-8}
\Delta E_0 = \int d\br\int d\brp K(|\br-\brp|) \bar{Q}_A(\br) \bar{Q}_B(\brp)
\end{eqnarray}
is the average (classical) electrostatic interaction energy between $A$ and $B$
while $\delta E_1$ and $\delta E_2$ contain the
correlated density fluctuations.
\begin{eqnarray}
\label{close-9}
\Delta E_1 &=& \frac{1}{2} \int_0^1 d\eta \int d\br\int d\brp K(|\br-\brp|)
[\bar{Q}_A(\br) \langle \delta \breve{Q}_{B+}(\brp)\rangle_\eta + \bar{Q}_B(\brp)
\langle \delta \breve{Q}_{A+}(\br)\rangle_\eta]\nonumber\\
\Delta E_2 &=&  \frac{1}{2} \int_0^1 d\eta \int d\br\int d\brp K(|\br-\brp|)
\langle \delta \breve{Q}_{A+}(\br)\delta \breve{Q}_{B+}(\brp)\rangle_\eta
\end{eqnarray}

Using the interaction picture with respect to $\breve{H}_{AB}$, each correlation
function in Eq. (\ref{close-9}) can be expressed as
\begin{eqnarray}
\label{close-10}
&&\langle \delta\breve{Q}_{A+}(\br)\delta\breve{Q}_{B+}(\brp)\rangle_\eta \nonumber\\
&=& \left\langle \breve{T} \delta\breve{Q}^I_{A+}(\br,t)\delta\breve{Q}^I_{B+}(\brp,t)
\mbox{exp}\left\{-i\sqrt{2}\int_{-\infty}^t d\tau \eta \breve{H}^I_{AB-}(\tau)\right\}
\right\rangle_0,
\end{eqnarray}
where the trace is with respect to the direct product
density-matrix of systems $A$ and $B$.
By expanding the exponential we can factorize the correlation function at each
order into a product of
density-correlation functions for system $A$ and $B$.
\begin{eqnarray}
\label{close-11}
&&\left\langle \delta\breve{Q}^I_{A\alpha_1}(\br,\tau_1)\cdots \delta\breve{Q}^I_{A\alpha_n}(\br,\tau_n)
\delta\breve{Q}^I_{B\alpha_{n+1}}(\br,\tau_{n+1})\cdots \delta\breve{Q}^I_{B\alpha_{n+m}}(\br,t_{n+m})
\right\rangle\nonumber\\
&=& \left\langle \delta\breve{Q}^I_{A\alpha_1}(\br,\tau_1)\cdots \delta\breve{Q}^I_{A\alpha_n}(\br,\tau_n)
\right\rangle_A
\left\langle \delta\breve{Q}^I_{B\alpha_{n+1}}(\br,\tau_{n+1})\cdots \delta\breve{Q}^I_{B\alpha_{n+m}}(\br,t_{n+m})\right\rangle_B.\nonumber\\
\end{eqnarray}
To lowest order in the interaction $K(|\br-\brp|)$, $\Delta E_2$ gives the well known
McLachlan's expression, $\Delta E_{MC}$, for the Van der Waals' energy\cite{mclachlan}
\begin{eqnarray}
\label{close-12}
\Delta E_{MC} &=&  \frac{1}{4}\int d\br \int d\brp \int d\br_1 \int d\brp_1\int dt \int dt_1K(|\br-\brp|)K(|\br_1-\brp_1|)\nonumber\\
&\times&\left[\frac{}{}R_{A++}(\br,t;\br_1,t_1)R_{B+-}(\brp,t;\brp_1,t_1)\right.\nonumber\\
&&\quad\quad\quad\quad\quad\quad+\left. R_{A+-}(\br,t;\br_1,t_1)R_{B++}(\brp,t;\brp_1,t_1)\frac{}{}\right]
\end{eqnarray}
where $R_{++}$ and $R_{+-}$ are the correlation and the response functions of
density-fluctuations, respectively.
\begin{eqnarray}
\label{close-13}
R_{X++}(\br,t;\brp,t_1) &=&\left\langle\breve{T}\delta \breve{Q}_{X+}(\br,t)\delta \breve{Q}_{X+}(\brp,t_1)
\right\rangle_X \nonumber\\
R_{X+-}(\br,t;\brp,t_1) &=&-i\left\langle\breve{T}\delta \breve{Q}_{X+}(\br,t)\delta \breve{Q}_{X-}(\brp,t_1)\right\rangle_X.
\end{eqnarray}
Using the Fourier transform $f(\omega)=\int dt e^{-i\omega t}f(t)$ and the fact that $R_{++}$ and $R_{+-}$ only depend on the difference of their time arguments, Eq. (\ref{close-12}) can be expressed in the frequency domain as
\begin{eqnarray}
\label{close-14}
\Delta E_{MC} &=&  \frac{1}{4}\int d\br \int d\brp \int d\br_1 \int d\brp_1\int \frac{d\omega}{2\pi}K(|\br-\brp|)K(|\br_1-\brp_1|)\nonumber\\
&\times&\left[\frac{}{}R_{A++}(\br,\br_1;\omega)R_{B+-}(\brp,\brp_1;\omega)\right.\nonumber\\
&&\quad\quad\quad\quad\quad\quad+\left. R_{A+-}(\br,\br_1;\omega)R_{B++}(\brp,\brp_1;\omega)\frac{}{}\right].
\end{eqnarray}

This is a general expression (to lowest order in the coupling) for the interaction energy in terms of the charge-density fluctuations ($R_{++}$) and response functions ($R_{+-}$) of both systems. Using the fluctuation-dissipation relation\cite{cohenPRL,up-shaulPRA}
\begin{eqnarray}
\label{close-15}
R_{X++}(\br,\brp,\omega)= 2\mbox{coth}\left(\frac{\beta\omega}{2}\right)\mbox{Im}R_{X+-}(\br,\brp,\omega),
\end{eqnarray}
where $\beta=1/k_BT$, the interaction energy can be
expressed solely in terms of the density-response $R_{+-}(\omega)$ of the two systems.
\begin{eqnarray}
\label{close-16}
\Delta E_{MC} &=&  \int d\br \int d\brp \int d\br_1 \int d\brp_1
\int \frac{d\omega}{4\pi}K(|\br-\brp|)K(|\br_1-\brp_1|)
\nonumber\\
&\times&\mbox{coth}\left(\frac{\beta\omega}{2}\right)
\left[\frac{}{}R_{A+-}(\br,\br_1;\omega)\mbox{Im}R_{B+-}(\brp,\brp_1;\omega)\right.\nonumber\\
&&\quad\quad\quad\quad\quad\quad+\left. R_{B+-}(\brp,\brp_1;\omega)\mbox{Im}R_{A+-}(\br,\br_1;\omega)
\frac{}{}\right].
\end{eqnarray}

When the two systems are well separated (compared to their sizes) we can expand the the coupling $K|\br-\brp|$ in multipoles around the position (charge centers) of the
two systems $\br^0_A$ and $\br^0_B$. $\hat{H}_{AB}$ can be
expressed in terms of multipoles of the two systems.
\begin{eqnarray}
\label{22april-1}
\hat{H}_{AB}&=& {\cal J}_{ij}\mu_{ai}\mu_{bj}
+{\cal K}_{ijk}[\mu_{ai}\Theta_{bjk}-\Theta_{aij}\mu_{bk}]+\cdots
\end{eqnarray}
where $\mu_{a}$ and $\Theta_{a}$ are dipole and quadruple operators for molecule $A$ and
indices $i,j,k$ denote the cartesian axis.
\begin{eqnarray}
\label{moments}
{\cal J}_{ij}&=& \left.\nabla_i\nabla^\prime_j
\frac{1}{|\br-\brp|}\right|_{{\br=\br^0_{A},{\brp=\brp^0_{B}}}}\nonumber\\
{\cal K}_{ijk}&=&\left. \frac{1}{3}\nabla_i\nabla^\prime_j\nabla^\prime_k \frac{1}{|\br-\brp|}
\right|_{{\br=\br^0_{A},{\brp=\brp^0_{B}}}}
\end{eqnarray}

The leading dipole-dipole term in Eq. (\ref{22april-1}),
which varies as $\sim 1/|\br-\brp|^3$ in Eq. (\ref{22april-1}),
dominates at large separations. Assuming for simplicity that the dipoles
of systems $A$ and $B$ are aligned along the same cartesian axis, the
matrix element ${\cal J}_{ij}$ in Eq. (\ref{moments}) reduces to a number
equal to $1/R^6$, where $R=|\br^0_A-\br^0_B|$ is the distance between the centers of the two dipoles. Equation (\ref{close-14}) now becomes
\begin{eqnarray}
\label{close-61a}
\Delta E_{MC}= \frac{1}{4R^6}\int \frac{d\omega}{2\pi}
\left[\alpha_{A++}(-\omega)\alpha_{B+-}(\omega)
+\alpha_{B++}(-\omega)\alpha_{A+-}(\omega)\right].\nonumber\\
\end{eqnarray}
Using the fluctuation-dissipation relation,  it can be recast as
\begin{eqnarray}
\label{close-61}
\Delta E_{MC}= \frac{1}{R^6}\int \frac{d\omega}{4\pi}\mbox{coth}\left(\frac{\beta\omega}{2}\right)
\left[\alpha_{A+-}(-\omega)\mbox{Im}\alpha_{B+-}(\omega)
+\alpha_{B+-}(-\omega)\mbox{Im}\alpha_{A+-}(\omega)\right].\nonumber\\
\end{eqnarray}
$R_{++}$ and $R_{+-}$ are now replaced by generalized polarizabilities $\alpha_{++}$
and $\alpha_{+-}$
\begin{eqnarray}
\label{close-20}
\alpha_{X++}(t-t_1) &=&\left\langle\breve{T} \breve{\mu}_{X+}(t)
\breve{\mu}_{X+}(t_1) \right\rangle_X \nonumber\\
&=& \alpha_X(t-t_1)+\alpha_{X}(t_1-t)\nonumber\\
\alpha_{X+-}(t-t_1) &=&-i\left\langle\breve{T} \breve{\mu}_{X+}(t)
\breve{\mu}_{X-}(t_1)\right\rangle_X\nonumber\\
&=& \theta(t-t_1)[\alpha_X(t-t_1)-\alpha_{X}(t_1-t)]
\end{eqnarray}
where $\alpha_X(t-t_1)=:\langle \hat{\mu}_X(t)\hat{\mu}_X(t_1)\rangle$ is the dipole correlation function of system $X$.

We can expand $\alpha_{++}$ and $\alpha_{+-}$ in terms of the eigenstates and
eigenvalues ($|a\rangle$,$\omega_a$) and ($|b\rangle$, $\omega_b$) of systems $A$ and $B$,
\begin{eqnarray}
\alpha_{A++}(t-\tp) & = & 2 \sum_{a\aap}P(a)|\mu_{a\aap}|^2 \mbox{cos}(\omega_{a\aap}(t-\tp))\nonumber\\
\label{close-21}\\
\alpha_{A+-}(t-\tp) & = & 2 \theta(t-\tp)\sum_{a\aap}P(a)
|\mu_{a\aap}|^2\mbox{sin}(\omega_{a\aap}(t-\tp))
\label{close-21a}
\end{eqnarray}
where $\mu_{a\aap}=\langle a|\mu|\aap \rangle$ and $P(a)=\mbox{e}^{-\beta E_a}$.
Corresponding expressions for system $B$ can be obtained by replacing indices $A$ and $a$
with $B$ and $b$ in Eqs. (\ref{close-21}) and (\ref{close-21a}). This gives,
\begin{eqnarray}
\alpha_{A++}(\omega) = \alpha_A^\prime(\omega),~~~~
\mbox{Im}\alpha_{A+-}(\omega) = \alpha_A^{\prime\prime}(\omega),~~~~
\end{eqnarray}
where $\alpha_A^\prime(\omega)$ and
$\alpha_A^{\prime\prime}(\omega)$ are the real and imaginary parts of the susceptibility,
\begin{eqnarray}
\label{close-25}
\alpha_A(\omega) = i\sum_{a\aap}P(a)|\mu_{a\aap}|^2
\left[\frac{1}{\omega-\omega_{a\aap}+i\eta}-\frac{1}{\omega+\omega_{a\aap}-i\eta}
\right]
\end{eqnarray}
with $\alpha^*_A(\omega)=\alpha_A(-\omega)$.

Using Eq. (\ref{close-25}), it is straightforward to show that
\begin{eqnarray}
\label{close-30}
\alpha_A^\prime(\omega) &=& \frac{1}{2}(1+\mbox{e}^{-\beta\omega})\alpha_A(\omega)\nonumber\\
\alpha_A^{\prime\prime}(\omega)&=&  \frac{1}{2}(1-\mbox{e}^{-\beta\omega})\alpha_A(\omega)
\end{eqnarray}
which gives the fluctuation-dissipation relation
\begin{eqnarray}
\label{close-40}
\alpha^\prime_A(\omega) = \mbox{coth}\left(\frac{\beta\omega}{2}\right) \alpha^{\prime\prime}_A(\omega).
\end{eqnarray}
Using Eq. (\ref{close-40}) in Eq. (\ref{close-21}) and taking inverse Fourier transform gives,
\begin{eqnarray}
\label{close-50}
\alpha_{A++}(\omega)= 2\mbox{coth}\left(\frac{\beta\omega}{2}\right)
\mbox{Im}\alpha_{A+-}(\omega)
\end{eqnarray}
The explicit expressions for $\alpha_{A++}$ and $\alpha_{A+-}$ are
\begin{eqnarray}
\label{close-60}
\alpha_{A++}(\omega)&=& \sum_{a\aap}P(a)|\mu_{a\aap}|^2(\delta(\omega+\omega_{a\aap})
+\delta(\omega-\omega_{a\aap}))
\nonumber\\
\alpha_{A+-}(\omega)&=& \sum_{a\aap}P(a)|\mu_{a\aap}|^2
\left[\frac{1}{\omega+\omega_{a\aap}+i\eta}-\frac{1}{\omega-\omega_{a\aap}+i\eta}\right].
\end{eqnarray}

Since $\mbox{Re}\alpha_{+-}(\omega)$ is symmetric while $\mbox{Im}\alpha_{+-}(\omega)$ is asymmetric in $\omega$, Eq. (\ref{close-61}) reduces to
\begin{eqnarray}
\label{close-62}
\Delta E_{MC}= \frac{1}{R^6}\int \frac{d\omega}{4\pi}\mbox{coth}\left(\frac{\beta\omega}{2}\right)
\mbox{Im}\left\{\alpha_{A+-}(\omega)\alpha_{B+-}(\omega)\right\}.
\end{eqnarray}
We can further simplify this expression by noting that $\mbox{Re}(\alpha_{++}(\omega)\alpha_{+-}(\omega))$ is
symmetric and contributes to the integral only at the pole of $\mbox{coth}(\beta\omega/2)$ for $\omega=0$
and this contribution vanishes if we take the principal part. Thus we get,
\begin{eqnarray}
\label{close-63}
\Delta E_{MC}= \frac{1}{R^6}\mbox{PP}\int \frac{d\omega}{4\pi i}\mbox{coth}\left(\frac{\beta\omega}{2}\right)
\alpha_{A+-}(\omega)\alpha_{B+-}(\omega)
\end{eqnarray}
where $ \mbox{PP}$ denotes the principal part. Thus interaction energy can be expressed solely in terms of the
response functions $\alpha_{+-}$ of the two systems. At high temperatures the integral in Eq. (\ref{close-63}) can be expanded in terms
of Matsubara frequencies defied by the poles of $\mbox{coth}(\beta\omega/2)$.
This gives
\begin{eqnarray}
\label{close-64}
\Delta E_{MC}= \frac{k_BT}{R^6} {\sum^{n=\infty}_{n=0}}^\prime \alpha_{A+-}(i\omega_n)\alpha_{B+-}(i\omega_n)
\end{eqnarray}
 where $\omega_n=2\pi n i/\beta$ and a $\;^\prime$ over the sum indictes a half contribution at the ploe
$\omega=0$. Eq. (\ref{close-64}) is the celebrated McLachlan expression \cite{mclachlan} for the
interaction energy of two coupled molecules with polarizabilities $\alpha_{A+-}$ and $\alpha_{B+-}$.

In Sec. \ref{observables} we shall follow the same procedure to compute different properties of interacting {\em open}
fermionic many-body systems.


\section{Superoperator algebra for fermion operators}\label{negf-fermion}

To combine the results of sections (\ref{superoperator}) and (\ref{closed-systems}) with many-body theory, we switch to second quantization. Here we consider fermionic systems. Bosonic systems are treated in Appendix \ref{boson}.
Electron creation (annihilation) operators, $\hat{\psi}(\hat{\psi}^\dag)$
satisfy the fermi anti-commutation relations.
\begin{equation}
\label{eq-1a} \{\hat{\psi}_{a},\hat{\psi}_{b}^{\dag }\}=\delta
_{ab},~\{\hat{\psi}_{a},\hat{\psi}_{b}\}=0,~\{\hat{\psi}_{a}^{\dag
},\hat{\psi}_{b}^{\dag }\}=0.
\end{equation}

The many-body density matrix can be represented in Fock space as $\hat{\rho}
=\sum_{MN}|M\rangle \langle N|$, where $|M\rangle $ and $|N\rangle $ are
many-body basis states with $m$ and $n$ electrons, respectively. These are
obtained by multiple operations of the Fermi creation operators on the
vacuum state
\begin{eqnarray}
\label{m-b-s}
|M\rangle &=&\hat{\psi}_{m}^{\dag }\cdots \hat{\psi}_{2}^{\dag }\hat{\psi}
_{1}^{\dag }|0\rangle,~~~
\langle M| = \langle 0|\hat{\psi}_{1}\hat{\psi}_{2}\cdots\hat{\psi}_{m}
\end{eqnarray}
$|M\rangle $ is antisymmetric with respect to the
interchange of any two particles, i.e., it must change sign when any
two out of the $m$ indices are interchanged on the right hand side of
Eq. (\ref{m-b-s}). This state can also be expressed as the direct
product of $m$ single particle states summed over all possible $m!$
permutations
\begin{equation}
|M\rangle =\sum_{i=1}^{m!}\frac{(-1)^{i}}{\sqrt{m!}}\hat{P}_{i}|m\rangle
\cdots |2\rangle |1\rangle
\end{equation}
where $|1\rangle =\hat{\psi}_{1}^{\dag }|0\rangle $, etc. are the single
particle states and $\hat{P}_{i}$ is the permutation operator which
generates $i$ successive permutations of two particles.

With each Hilbert space operator ${\hat{\psi}}$ we associate a pair of superoperators
in Liouville space, ${\breve{\psi}}_{L}$, and ${\breve{\psi}}_{R}$, defined
through their action on a Liouville space vector \cite{manfred}
\begin{eqnarray}
{\breve{\psi}}_{L}|M\rangle \langle N| &\equiv &{\hat{\psi}}|M\rangle \langle
N|,~~{\breve{\psi}^{\dagger }}_{L}|M\rangle \langle N|\equiv {\hat{\psi}%
^{\dagger }}|M\rangle \langle N|  \nonumber \\
{\breve{\psi}}_{R}|M\rangle \langle N| &\equiv &(-1)^{m-n}|M\rangle \langle N|{%
\hat{\psi}}\nonumber\\
{\breve{\psi}^{\dagger }}_{R}|M\rangle \langle N|&\equiv & (-1)^{m-n+1}|M\rangle
\langle N|{\hat{\psi}^{\dagger }}\label{eq-3aa}.
\end{eqnarray}
The definitions for the "right" operators differ from
Eq. (\ref{liouv-operator}) by the $(-1)$ factors.
Note that ${\breve{\psi}^{\dagger }}_{R}$ is
not the hermitian conjugate of ${\breve{\psi}}_{R}$ as will be shown below.
These are introduced in order to account more conveniently for the Fermi statistics.
These factors are not required for bosons where the many-body state
is symmetric with respect to particle exchange (Appendix \ref{boson})
and we can simply use the single-body definitions, Eq. (\ref{liouv-operator}).
{\em The superoperators defined in Eq. (\ref{eq-3aa}) satisfy the same anti-commutation
relations as their Hilbert space counterparts} [Eq. (\ref{eq-1a})]
\begin{eqnarray}
\{{\breve{\psi}}_{m\alpha },{\breve{\psi}^{\dagger }}_{n \beta
}\} &=&\delta _{mn}\delta _{\alpha \beta },  \nonumber  \label{eq-supercommu} \\
\{\breve{\psi}_{m\alpha },{\breve{\psi}}_{n\beta }\} &=&0,~~~~\{{\breve{\psi}^{\dagger }}%
_{m\alpha},{\breve{\psi}^{\dagger }}_{n\beta }\}=0.
\end{eqnarray}%

The fermion superoperators in the PTSA representation are defined
by Eqs. (\ref{trans}).
Substituting (\ref{eq-supercommu}) in Eqs. (\ref{pm-1}) and (\ref{pm}), we obtain for
fermion operators ($\breve{\chi},\breve{\zeta}=\breve{\psi},\breve{\psi}^\dag$)
\begin{eqnarray}
\label{fermi-product}
(\hat{\chi}\hat{\zeta})_{+} &=&\frac{1}{\sqrt{2}}\left[
\breve{\chi}_{+}\breve{\zeta}_{-}
+\breve{\chi}_{-}\breve{\zeta}_{+}+(1-\delta _{\chi,\zeta})
\right],
(\hat{\chi}\hat{\zeta})_{-} =\frac{1}{\sqrt{2}}\left[
\breve{\chi}_{+}\breve{\zeta}_{+}
-\breve{\zeta}_{-}\breve{\chi}_{-}\right]
\end{eqnarray}
where $\delta _{\chi,\zeta}$ is unity if $\chi$ and $\zeta$ are same type (creation or
annihilation) operators and zero otherwise.
Equations \ref{eq-supercommu} hold in both PTBK ($\alpha,\beta=L,R$) and PTSA ($\alpha,\beta=+,-$) representations.

From Eqs. (\ref{eq-3aa}), we obtain
\begin{eqnarray}
\label{ccc-1}
\langle \breve{\psi}_R\rangle&=& \mbox{Tr}\{\breve{\psi}_R\hat{\rho}\}
= \mbox{Tr}\{\sum_{MN}\breve{\psi}_R|M\rangle\langle N|\}\nonumber\\
&\equiv& \mbox{Tr}\{\sum_{MN}(-1)^{m-n}|M\rangle\langle N|\hat{\psi}\}\nonumber\\
&=& \sum_{M}\langle M+1|M\rangle= \mbox{Tr}\{\sum_{MN}|M\rangle\langle N+1|\} \nonumber\\
&=& \mbox{Tr}\{\sum_{MN}|M\rangle\langle N|\hat{\psi}\} = \mbox{Tr}\{\hat{\psi}\hat{\rho}\}=\langle \hat{\psi}\rangle
\end{eqnarray}
where in going from second to the third line the sum over $N$ can be done using the fact that trace is the sum of the diagonal elements of the matrix $|M\langle\rangle N+1|$.
Using the same arguments we can write
\begin{eqnarray}
\label{ccc-2}
\langle \breve{\psi}_R^\dag\rangle&=& \mbox{Tr}\{\breve{\psi}_R^\dag\hat{\rho}\}
= \mbox{Tr}\{\sum_{MN}\breve{\psi}_R^\dag|M\rangle\langle N|\}\nonumber\\
&\equiv& \mbox{Tr}\{\sum_{MN}(-1)^{m-n+1}|M\rangle\langle N|\hat{\psi}^\dag\}\nonumber\\
&=&\mbox{Tr}\{\sum_{MN}(-1)^{m-n+1}|M\rangle\langle N-1|\}\nonumber\\
&=& - \sum_{M}\langle M-1|M\rangle= -\mbox{Tr}\{\sum_{MN}|M\rangle\langle N-1|\}\nonumber\\
&=& -\mbox{Tr}\{\hat{\psi}^\dag\hat{\rho}\}=-\langle \hat{\psi}^\dag\rangle.
\end{eqnarray}
Similarly it can be shown that, $\langle \breve{\psi}_L\rangle=\langle \hat{\psi}\rangle$ and
$\langle \breve{\psi}_L^\dag\rangle=\langle \hat{\psi}^\dag\rangle$.

Using Eqs. (\ref{ccc-1}) and (\ref{ccc-2}) we get
\begin{eqnarray}
\label{nn-11}
\langle \breve{\psi}_-\rangle &=& \frac{1}{\sqrt{2}}\left[\langle  \breve{\psi}_L \rangle
-\langle  \breve{\psi}_R \rangle \right]=0\nonumber\\
\langle \breve{\psi}_+\rangle &=& \frac{1}{\sqrt{2}}\left[\langle  \breve{\psi}_L \rangle
+\langle  \breve{\psi}_R \rangle \right]
\equiv\sqrt{2} \langle \hat{\psi} \rangle
\end{eqnarray}
and
\begin{eqnarray}
\label{nn-22}
\langle \breve{\psi}_-^\dag\rangle &=& \frac{1}{\sqrt{2}}\left[\langle  \breve{\psi}_L^\dag \rangle
-\langle  \breve{\psi}_R^\dag \rangle \right]
\equiv  \sqrt{2}\langle\hat{\psi}^\dag \rangle\nonumber\\
\langle \breve{\psi}_+^\dag\rangle &=& \frac{1}{\sqrt{2}}\left[\langle  \breve{\psi}_L^\dag \rangle
+\langle  \breve{\psi}_R^\dag \rangle \right]=0.
\end{eqnarray}
Thus $\breve{\psi}$ satisfies the relations Eqs. (\ref{liu-hil}) while $\breve{\psi}^\dag$ does
not. Note however that any $\breve{A}_\alpha$ of the form of a product of Fermi superoperators which contains
equal number of $\breve{\psi}$ and $\breve{\psi}^\dag$ always satisfies Eqs. (\ref{liu-hil}), since
it behaves like a boson operator$^2$. \footnotetext[2]{This is true for any operator containing product of even number of ordinary fermi operators.}

Wick's theorem, Eq. (\ref{expectation}), for products of fermion operators
accounts for the antisymmetry of the many-body state with respect
to the permutation of any two particles (Appendix \ref{wick}).
For fermion superoperators this reads,
\begin{eqnarray}
\langle \breve{\chi}_{1\alpha _{1}}\breve{\chi}_{2\alpha _{2}}\cdots \breve{\chi}
_{(n-1)\alpha _{n-1}}\breve{\chi}_{n\alpha _{n}}\rangle
=\sum_{P}(-1)^{p}\langle \breve{\chi}_{1\alpha _{1}}\breve{\chi}_{2\alpha
_{2}}\rangle \cdots \langle \breve{\chi}_{(n-1)\alpha _{n-1}}\breve{\chi}_{n\alpha
_{n}}\rangle
\end{eqnarray}
where the index $p$ denotes the number of permutations required to arrive at the desired
pairing.

\section{SNGF for fermions}
\label{gf-nn}

Conventional Hilbert-space CTPL is based on four
basic closed time path Green's
functions  \cite{keldysh,chou,hao,mills,lifshitz,eijck}).
The time contour in CTPL is defined by parameterizing the
real time in terms of a loop-time parameter $\tau $. A two-time
Green's function, $G(\tau _{1},\tau _{2})$,
is defined on the contour. Depending on which branch of the contour
$\tau _{1}$ and $\tau _{2}$ lie, i.e., by ordering the two
time-points on the contour, one obtains four Green's functions
($G^{>}$, $G^{<}$, $G^{T}$ and $G^{\tilde{T}}$), each describing
a distinct physical process. By formulating
Green's function theory in Liouville space, we find that
$G^{>}$ and $G^{<}$ are related to transport while $G^{T}$ and $G^{
\tilde{T}}$ represent Fock space coherences\cite{up-shaul-prb1}.

We now have all the ingredients necessary to introduce the SNGF
\begin{equation}
\label{liouville-green} G_{\alpha\beta}^{m,n}(t,{t^{\prime
}}):= -i\langle \breve{T}\breve{\psi} _{m\alpha }(t)\breve{\psi}
_{n\beta }^{\dag }({t^{\prime }})\rangle,
\end{equation}
where the indices $m,n$ represent the single particle degrees of freedom, and
all operators are in Heisenberg picture
\begin{eqnarray}
\breve{\psi}_{m\alpha}(t)= \mbox{e}^{i\sqrt{2}\breve{H}_-t}\breve{\psi}_{m\alpha}\mbox{e}^{-i\sqrt{2}\breve{H}_-t}\nonumber\\
\breve{\psi}^\dag_{m\alpha}(t)= \mbox{e}^{i\sqrt{2}\breve{H}_-t}\breve{\psi}^\dag_{m\alpha}\mbox{e}^{-i\sqrt{2}\breve{H}_-t}.
\end{eqnarray}
The trace $\langle \cdots \rangle $ in Eq. (\ref{liouville-green}) is with respect to the density
matrix of the interacting system (Eq. (\ref{rho-1}) with $\hat{H}$ given by Eq. (\ref{close-1})).
$\breve{T}$ is a key formal tool in this approach  which
allows to derive compact expressions for all Green's
functions and observables. Each permutation of Fermi superoperators, $\breve{\psi}
_{\alpha }$ and $\breve{\psi} _{\alpha }^{\dag }$, necessary to
achieve the desired time ordering brings in a factor of $(-1)$. This
sign convention associated with time ordering is essential for deriving the Dyson
equation for the Green's function, Eq. (\ref{liouville-green}).

In Appendix \ref{H-L-GF} we recast these Green's functions in Hilbert
space. In Appendix \ref{H-L-GF-2}, we show that $G_{--}$, $G_{++}$
and $G_{-+}$ coincide with the retarded, the advanced and the correlation
Green's functions, respectively.
\begin{eqnarray}
G^{mn}_{--}(t,{t^{\prime }}) &=&-i\theta (t-{t^{\prime }}%
)\langle \{\hat{\psi}_{m}(t),\hat{\psi}_{n}^{\dag }({t^{\prime
}})\}\rangle
\label{g--1} \\
G^{mn}_{++}(t,{t^{\prime }}) &=& i\theta ({t^{\prime }}%
-t)\langle \{\hat{\psi}_{m}(t),\hat{\psi}_{n}^{\dag }({t^{\prime
}})\}\rangle
\label{g++1} \\
G^{mn}_{-+}(t) &=&-i\langle \lbrack \hat{\psi}
_{m}(t),\hat{\psi}_{n}^{\dag }({t^{\prime }})]\rangle ]  \label{g-+1} \\
G_{+-}^{mn} &=&0\label{g+-1}.
\end{eqnarray}

Consider two interacting fermion systems $A$ and $B$ described by
the Hamiltonian in Eq. (\ref{close-1}) where the coupling is now given by
\begin{equation}
\label{hab}
\hat{H}_{AB}=\sum_{ab}J_{ab}{\hat{\psi}^{\dagger
}}_{a}{\hat{\psi}} _{b}+h.c.\;.
\end{equation}
The indices $a$($b$) run over the orbital and/or spin degrees of freedom
associated with system $A$($B$). Such coupling appears in a wide
variety of physical systems \cite{biological}.
We now show how properties of the joint system may be expressed in terms of
one-particle SNGF of the individual systems.

At $t=-\infty$ we assume that the total density matrix is given by a
direct product of density matrices of the individual systems,
$\hat{\rho}(-\infty)=\hat{\rho}_{A}\oplus\hat{\rho}_{B}$. The
density matrix of the interacting system can then be constructed
by turning on the interaction $H_{AB}$ adiabatically.
\begin{equation}
\label{dm} \hat{\rho}(t)=\breve{U}(t,-\infty)\hat{\rho}(-\infty)
\end{equation}
where
\begin{equation}
\breve{U}(t,-\infty)=\breve{T}\mbox{exp}\left[
-i\sqrt{2}\int_{-\infty}^{t}d \tau {\breve{H}^I_{AB-}}(\tau )\right].
\end{equation}
$\breve{H}^I_{AB-}(t)$ is the superoperator corresponding to the
Hamiltonian $\hat{H}_{AB}$ in the interaction representation.
\begin{equation}
\breve{H}^I_{AB-}(t)=\breve{U}^\dag(t,0)\breve{H}_{AB}\breve{U}(t,0)
\end{equation}
where $\breve{U}_0(t,0)=\theta(t)e^{-i\breve{H}_{0-}t}$ and $\breve{H}_{0-}$ is the
superoperator corresponding to Hamiltonian,
$\hat{H}_{0}=\hat{H}-\hat{H}_{AB}$. Using the interaction
representation of the density matrix, Eq. (\ref{dm}), the SNGF assumes the form,
\begin{equation}
\label{green-interaction}
G_{\alpha \beta }^{m,n}(t,{t^{\prime }})=-i\left\langle \breve{T}{\breve{\psi}}
_{m\alpha }(t){\breve{\psi}^{\dagger }}_{n\beta }({t^{\prime
}})\mbox{exp}\left[ -i\sqrt{2}
\int_{-\infty}^{t}d\tau \breve{H}^I_{AB-}(\tau
)\right] \right\rangle_0
\end{equation}
where the trace is now taken with respect to non-interacting density matrix $\rho(-\infty)=\rho_A\oplus\rho_B$.

Equation (\ref{green-interaction}) is a convenient starting point
for computing the
Green's functions perturbatively in $\breve{H}_{AB-}$.
Each term can be expressed as a
product of SNGF of the individual systems $A$ and $B$.
The Green's functions of
individual systems can be calculated perturbatively by solving the matrix Dyson equation derived in
section \ref{sec-dyson}.

\section{Superoperator Green's-function expressions for observables}
\label{observables}

We consider two interacting
systems $A$ and $B$ described by the Hamiltonian, Eqs.
(\ref{close-1}) and (\ref{hab}).
For observables such as the charge density
described by operator $\hat{A}$ which contain an equal number
of creation and annihilation operators, ${\hat{\psi}}$ and
${\hat{\psi}^{\dagger }}$ we have $\langle \hat{A}_{L}\rangle
=\langle \hat{A}_{R}\rangle \equiv \langle \hat{A}\rangle $ and
$\langle \hat{A} _{-}\rangle =0$. Using a perturbative expansion
in $\hat{H}_{AB}$ we can express the various observables in
terms of Green's functions of the individual systems.

\subsection{The interaction energy}

The interaction energy $\Delta E$ is defined by Eq. (\ref{close-3}) where $\hat{H}_{AB}$ is given by Eq. (\ref{hab}).
We follow the procedure outlined in Sec. \ref{closed-systems} to compute $\Delta E$.
We introduce a switching parameter ($\eta $) as in Eq. (\ref{eta-hamil}).
For $\eta =0$, $\hat{H}_{\eta }$ describes the noninteracting
subsystems $A$ and $B$, whereas for $\eta =1$, we get the full
Hamiltonian $\hat{H}_{\lambda }=\hat{H}$. $\eta$ is a
convenient bookkeeping device. Using the Helmann-Feynman theorem,
interaction energy can be expressed as in Eq. (\ref{close-4}).

In the interaction representation Eq. (\ref{close-4}) reads
\begin{eqnarray}
\Delta E = \frac{1}{\sqrt{2}}\int_{0}^{1}d\eta
\left\langle \breve{T}\breve{H}_{AB+}^I(t)\mbox{exp}
\left\{-i\eta\sqrt{2}\int_{-\infty}^t d\tau \breve{H}^I_{AB-}(\tau) \right\}\right\rangle _{\eta}.
\end{eqnarray}
By expanding the exponential we can compute the interaction energy perturbatively in the
coupling $J_{AB}$, Eq. (\ref{hab}). To lowest order we get,
\begin{eqnarray}
\label{int-energy-j2}
\Delta E &=&\frac{1}{2}\mbox{Im}\sum_{aba^{\prime }b^{\prime
}}J_{ab}J_{a^{\prime }b^{\prime }}\int \frac{d\omega }{2\pi }
\left[ G^{a^{\prime }a}_{++}(\omega )G^{bb^{\prime }}_{-+}(\omega
)+G^{a^{\prime }a}_{-+}(\omega )G^{bb^{\prime }}_{--}(\omega )\right] .
\end{eqnarray}
This extends the McLachlan's expression, Eq. (\ref{close-61a}), for Van der Waals forces
\cite{mclachlan,up-shaulPRA,stone,cohenPRL} to open fermionic systems.

We next assume that system $B$ is infinitely
large and therefore unaffected by the coupling with the
small system $A$, and remains at thermodynamic equilibrium at all times.
This corresponds,
$e.$$g.$ to a molecule ($A$) chemisorbed on a metal surface ($B$).
We further neglect electron-electron ($e-e$) interactions in system $B$ and set
$H_{B}=\sum_{b}\epsilon_b\psi _{b}^{\dag }\psi _{b}$. Its Green's functions are
then given by
\begin{eqnarray}
G_{LR}^{b{b^{\prime }}}(t,{t^{\prime }}) &=&i\delta _{b{b^{\prime }}}f_{b}%
\mbox{e}^{-i\epsilon _{b}(t-{t^{\prime }})}  \label{zeroth-green-func} \\
G_{RL}^{b{b^{\prime }}}(t,{t^{\prime }}) &=&i\delta _{b{b^{\prime }}%
}(1-f_{b})\mbox{e}^{-i\epsilon _{b}(t-{t^{\prime }})} \\
G^{b{b^{\prime }}}_{--}(t,{t^{\prime }}) &=&-i\delta _{b{b^{\prime }}}\theta
(t-{t^{\prime }})\mbox{e}^{-i\epsilon _{b}(t-{t^{\prime }})}\label{last}
\end{eqnarray}
where $f_b=[1+\exp\{\epsilon_b-\epsilon_F\}]$ is the Fermi
function and $\epsilon_b$ is the energy of an orbital $(b)$ of system $B$.

Upon substituting in Eq. (\ref{int-energy-j2}) we get
\begin{eqnarray}
\label{limit-int-energy-j2}
\Delta E &=& \frac{1}{\sqrt{2}}\sum_{aa^\prime b} J_{ab}J_{a^\prime b}^*
\left[\mbox{Im}
\int \frac{d\omega}{2\pi} \frac{G_{-+}^{{a^\prime} a}(\omega)}{
\omega-\epsilon_b+i0^+}
+\mbox{Re}
[1-2f(\epsilon_b)]G_{++}^{{a^\prime} a}(\epsilon_b)\right].
\end{eqnarray}
This expression depends explicitly on the SNGF of the system $A$.
Electron-electron and electron-phonon interactions can be included
perturbatively using the matrix Dyson equations derived in Sec. \ref{sec-dyson}

\subsection{charge-density redistribution }

We now turn to a different observable, the fluctuations in the charge density $\delta n_{A}$ of system $A$ due to its interaction with $B$. The local charge density at
point $\br_{a}$ is given by the diagonal density-matrix element of molecule $A$,
\begin{eqnarray}
n_{A}(\br_{a})= \left.\rho _{A}(\br_{a},\br_{a}^{\prime })\right\vert
_{\br_{a}^{\prime}=\br_{a}}
\end{eqnarray}
where
\begin{eqnarray}
\label{dm-sys-A}
\hat{\rho}_A(\br_a,\br_a^\prime)&=&\langle \hat{\psi}^\dag(r_a) \hat{\psi}(\br^\prime_a)\rangle\nonumber\\
&\equiv& iG_{LR}(\br_a,t;\br_a^\prime,t).
\end{eqnarray}
is the density-matrix of system $A$.
The charge density fluctuation of system $A$ at point $r_a$ is
\begin{eqnarray}
\delta n_{A}(\br_{a})&=&\left.\delta\rho(\br_a,\br_a^\prime)\right|_{\br_a=\br_a^\prime}\nonumber\\
&=& \left|\rho_{A}(\br_a,\br_a^\prime)-\rho_{0A}(\br_a,\br_a^\prime)\right|_{\br_a=\br_a^\prime}
\end{eqnarray}
where $\delta\rho_A$ is the fluctuation in $\rho_A$ and
$\rho_{0A}\equiv\langle {\hat{\psi}^{\dagger }}_{A}(\br_{a}){\hat{\psi}}
_{A}(\br_{a}^\prime)\rangle _{0}$ is the density matrix of the isolated molecule $A$.

We shall compute the change in the density matrix of molecule $A$, $\delta
\rho_{A}(\br_{a},\br_{a}^{\prime })$ perturbatively in $J_{ab}$. To that end, using
Eqs. (\ref{green-interaction}) and (\ref{dm-sys-A}), we recast the density matrix
of system $A$ in the interaction picture
\begin{eqnarray}
\label{nmn-aa}
\rho_A(\br_a,\br_a^\dag) = -i\left\langle \breve{T} \breve{\psi}_L(\br_a^\prime,t)\breve{\psi}_R^\dag(\br_a,t)
\mbox{exp}\left[-i\sqrt{2}\int_{-\infty}^t d\tau \breve{H}_{AB-}^I(\tau) \right]\right\rangle_0
\end{eqnarray}
By expanding the exponential we can compute $\rho_A$ perturbatively in the
coupling between the two systems, Eq.(\ref{hab}). The first term in the expansion is
simply $\rho_{0A}$ and the second term (which is second
order in the coupling) gives the fluctuation of the density matrix.
\begin{eqnarray}
\label{charge-density}
&&\delta \rho _{A}(\br_{a},\br_{a}^{\prime })=-i\int \frac{d\omega }{%
8\pi }\int d{\bf r} J(\br_{a1}-\br_{b1})J^{\ast }(\br_{a2}-\br_{b2}) \nonumber
\label{charge-density-matrix} \\
&&\left[ G_{-+}(\omega ;\br_{a}^{\prime },\br_{a1})G_{++}(\omega
;r_{a2},\br_{a})G_{++}(\omega ;\br_{b1},\br_{b2})\right.   \nonumber \\
&+&\left.G_{-+}(\omega ;\br_{a2},\br_{a})G_{--}(\omega ;\br_{a}^{\prime
},\br_{a1})G_{--}(\omega ;\br_{b1},\br_{b2})\right.   \nonumber \\
&+&\left. G_{-+}(\omega ;\br_{b1},\br_{b2})G_{++}(\omega
;\br_{a2},\br_{a})G_{--}(\omega ;\br_{a}^{\prime },\br_{a1})\right]
\end{eqnarray}
where $d{\bf r}=d\br_{a1}d\br_{b1}d\br_{a2}d\br_{b2}$. Integrating Eq.
(\ref{charge-density}) over the region $A$ gives the
total change in the number of electrons, $\delta N_{A}$.
\begin{eqnarray}
\delta N_A = \int d\br_a \delta n_A(\br_a).
\end{eqnarray}
Note that $\delta N_{A}$ need not be an integer since the two systems can be entangled;
an electron can be delocalized across a sub-system, giving them a partial charge.
Stated differently, the system can have {\em Fock space coherences}. Generally, a many-body state $\Psi(N)$ of the
joint system with $N$ electrons can be expressed as combinations of
$\Psi_A(n)$ and $\Psi_B(N-n)$ as
\begin{equation}
\Psi(N) = \sum_{n,p,q}C^{n}_{pq}\Psi_{A}^{(n,p)}\Psi_{B}^{(N-n,q)}
\end{equation}
where $p,q$ represent various many-body states of $A$ and $B$ with $n$ and
$N-n$ electrons. This entangled many-body state may not be represented by an ensemble of
states with different $n$, as is commonly done in ensemble
DFT \cite{dft1,dft2,morrel-cohen,parr-yang,mermin}.

As we did for the interaction energy, we can derive a closed expression
for the change in the density matrix of system $A$, in the limiting case when system
$B$ is large and made of noninteracting electrons. Denoting the
single particle (orbital) wavefunctions of system $B$ by
$\varphi(r_{b})$, we get
\begin{eqnarray}
\label{eq-101}
G_{++}(\omega;\br_{b},\br_{b}^\prime)&=&\sum_b \frac{\varphi_b(\br_{b})
\varphi_b(\br_{b}^\prime)}{\omega-\epsilon_b-s\eta} \\
G_{-+}(\omega;\br_{b},\br_{b})&=&-i\sum_b \varphi_b(\br_{b})
\varphi_b(\br_{b}^\prime)[1-2f_b]\delta(\omega-\epsilon_b).
\end{eqnarray}
By substituting Eq. (\ref{eq-101}) in (\ref{charge-density-matrix}),
we obtain,
\begin{eqnarray}
\label{eq-102}
&&\delta\rho(\br_a,\br_a^\prime)=-\frac{1}{4}\sum_b\int\int
d\br_{a1}d\br_{a2}I_b(\br_{a1},\br_{a2})  \nonumber \\
&&\left[\frac{}{}(1-2f_b)G_{++}(\epsilon_b;\br_{a2},\br_{a1})
G_{--}(\epsilon_b;\br_{a}^\prime,\br_{a1})\right.  \nonumber \\
&+&\left.i\int\frac{d\omega}{2\pi}\left\{ \frac{G_{-+}(\omega;\br_a^%
\prime,\br_{a1}) G_{++}(\omega;\br_{a2},\br_a)} {\omega-\epsilon_b-i\eta}
\right.\right.  \nonumber \\
&+&\left.\left. \frac{G_{-+}(\omega;\br_{a2},\br_{a})G_{--}(\omega;\br_{a}^%
\prime,\br_{a2})} {\omega-\epsilon_b+i\eta} \right\}\right]
\end{eqnarray}
where
\begin{eqnarray}
I_b(\br_{a1},\br_{a2})&=&\int\int d\br_{b1} d\br_{b2} \varphi_b(\br_{b1})\varphi_b(\br_{b2}^\prime)
J(\br_{a1}-\br_{b1})J^*(\br_{a2}-\br_{b2}).
\end{eqnarray}

\subsection{The current in a junction}\label{observables-current}

The current is defined as the expectation value of the total electron flux from system $A$ to $B$.  The current operator is given
by as the rate of change of number of electrons in subsystem $A$.
\begin{eqnarray}
\hat{I}:=\frac{d}{dt}\hat{\psi} _{a}^{\dag }\hat{\psi} _{a}=i\left[ \hat{H}_{AB}
(t),\hat{\psi} _{a}^{\dag }\hat{\psi} _{a}\right].  \label{eq-1}
\end{eqnarray}
Substituting Eq. (\ref{hab}) we obtain
\begin{equation}
\hat{I}=ie\left( J_{ba}\psi _{b}^{\dag }\psi _{a}-J_{ab}\psi _{a}^{\dag
}\psi _{b}\right) .
\end{equation}
The current is given by the expectation value of $\hat{I}$ with respect
to the total (interacting) density matrix of the system.
\begin{eqnarray}
I(t)&=&-2eJ_{ab}\mbox{Im}\langle \psi _{b}^{\dag }\psi
_{a}\rangle=
2eJ_{ab}\mbox{Re} G_{LR}^{ab}(t,t),\label{current-1}
\end{eqnarray}
where in second equality we have used Eq.  (\ref{glr}).

As explained in Sec. \ref{negf-fermion}, we can expand the Green's
function in Eq. (\ref{current-1}) perturbatively in the coupling
between the two sub-systems. To lowest order this gives
\begin{eqnarray}
I(t)& =& e\sum_{ab{a^{\prime }}{b^{\prime }}}J_{ab}J_{{a^{\prime }}{b^{\prime }
}}^{\ast }\int d\tau \left[ G_{RL}^{{a^{\prime }}a}(\tau ,t)G_{LR}^{b{
b^{\prime }}}(t,\tau )
 - G_{RL}^{b{b^{\prime }}}(\tau ,t)G_{LR}^{{a^{\prime }}a}(t,\tau )
\right]  \label{eq-final-2-y}
\end{eqnarray}
When the couplings $J_{ab}$ are real, the current can also be recast in
simple form in terms of the PTSA $G^{--}$ and $G^{-+}$ (Eq.
(\ref{current-pp-mp})).
We note that the total current only depends on $G_{LR}$ and $G_{RL}$;
$G_{LL}$ and $G_{RR}$ do not contribute.
This can be understood naturally using the density matrix as depicted in
the double sided Feynman diagram as shown in Fig. 2.
$G_{LR}$ and $G_{RL}$ represent electron transfer between systems
($|N\rangle\langle N|$ changes to $|N+ 1\rangle\langle N+ 1|$ or
$|N-1\rangle\langle N-1|$)
while $G_{LL}$ and $G_{RR}$ only represent Fock space coherences with
no change of the number of electrons ($|N\rangle\langle N|$ remains unchanged at the end).
\begin{figure}[h]
\centering
\rotatebox{0}{\scalebox{.8}{\includegraphics{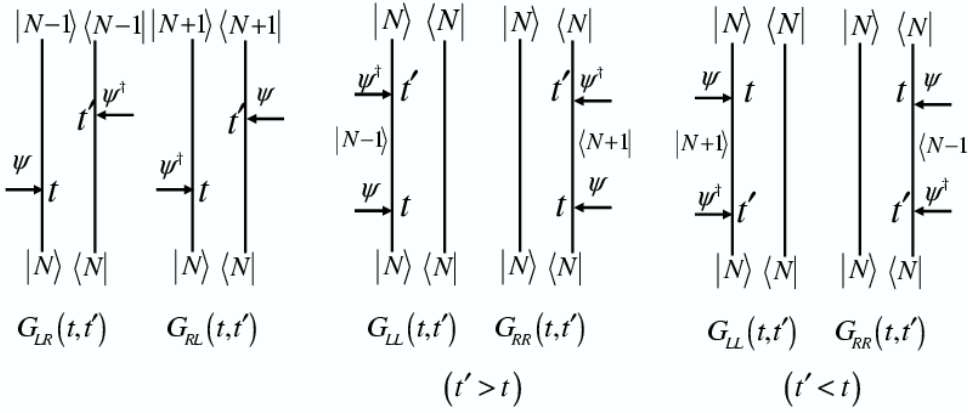}}}\newline
\caption{\em Double sided Feynman diagrams for the four superoperator Green's functions.
$|N\rangle$ represents a many-body state with $N$ electrons.  $G_{LR}$
and $G_{RL}$ correspond to the transition from $|N\rangle$ to $|N-1\rangle$
and  $|N\rangle$ to $|N+1\rangle$ state, respectively. $G_{LL}$
and $G_{RR}$ represent coherence between the two states during time
interval $t$ to $t^\prime$.
}
\label{fig3}
\end{figure}

In the long time limit, the system reaches a steady state where the
current becomes independent on time and the Green's functions only
depend on the difference of their time arguments.  Using Fourier
transformation of $G_{\alpha\beta}(E) = \int dt G_{\alpha\beta}(t)\e^{i E t}$, Eq.
(\ref{eq-final-2-y}) can be recast in the frequency domain as
\begin{eqnarray}
I& =& e\sum_{ab{a^{\prime }}{b^{\prime }}}J_{ab}J_{{a^{\prime }}{b^{\prime }
}}^{\ast }\int \frac{dE}{2\pi} \left[ G_{RL}^{{a^{\prime }}a}(E)G_{LR}^{b{
b^{\prime }}}(E)
- G_{RL}^{b{b^{\prime }}}(E)G_{LR}^{{a^{\prime }}a}(E)%
\right].  \label{eq-final-2-yy}
\end{eqnarray}

As done previously, [Eqs. (\ref{zeroth-green-func})-(\ref{last}), we next assume that sub-system $B$ is a thermodynamically large, free electrons system. in this case, the current can be calculated non-perturbatively to all orders in $J$.
This is derived in Appendix \ref{curr-exact}.
Equation (\ref{eq-final-2-yy}) can then be replaced by Eq. (\ref{eq-final-2-new})
\begin{eqnarray}
\label{nn-33}
I=e\sum_{a{a^{\prime}}}\int \frac{dE}{2\pi} \left[{%
\Sigma^{a{a^{\prime}}}_{LR}}(E)G_{RL}^{a{a^{\prime}}}(E)-{\Sigma^{a{
a^{\prime}}}_{RL}}(E)G_{LR}^{a{a^{\prime}}}(E) \right]
\end{eqnarray}
where $\Sigma_{\alpha\beta}^{aa^\prime}, \alpha,\beta=L,R$ are the elements
of the self-energy matrix for sub-system $A$ due to interaction with $B$.
\begin{eqnarray}
\Sigma^{a{a^{\prime}}}_{LR}(E)&=&i\Gamma^A_{aa^\prime}f_B(E),~~
\Sigma^{a{a^{\prime}}}_{RL}(E)=-i\Gamma^A_{aa^\prime}[1-f_B(E)]\\
\Sigma^{a{a^{\prime}}}_{LL}(E)&=& -\frac{i}{2}\Gamma^A_{aa^\prime}[1-2f_B(E)],~~
\Sigma^{a{a^{\prime}}}_{RR}(E)= \frac{i}{2}\Gamma^A_{aa^\prime}[1-2f_B(E)]
\end{eqnarray}
with $\Gamma^A_{aa^\prime}=
2\pi\sum_{bb^\prime}J_{ab}J^*_{a^\prime b^\prime}\sigma_B$, where $\sigma_B$
is the density of states (assumed to be independent of energy) of sub-system $B$.
Unlike Eq. (\ref{eq-final-2-yy}), the Green's
functions in Eq. (\ref{nn-33}) now contain coupling to the leads to all orders through
the self-energies. These can be evaluated by solving the matrix Dyson equation
as given in Sec. \ref{sec-dyson}.
Under certain
approximations, Eq. (\ref{nn-33}) reduces to the LB form, Eq. (\ref{nn-44}).
This is shown in Eqs. (\ref{new-lb-form})-(\ref{new-s}).


\subsection{Fluorescence induced by an electron or a hole transfer}

Consider an electron transferred to a molecule attached to two
leads (a molecular wire) or by an STM probe
by applying an external potential. This electron may decay radiatively to
one of the lower energy states by emitting a photon, before finally
exiting the molecule. We denote this optical signal current-induced-fluorescence (CIF).
The reverse process can
occur when an electron leaves the molecule, creating a
hole. An electron in one of the higher lying bound states can decay
radiatively to combine with this hole. Thus the CIF signal can come
either from a positively or from a negatively charged molecule where the
charge is created by the interaction with external leads. This is
different from ordinary (laser-induced) fluorescence (LIF)
\cite{shaul-book} where an {\em electron-hole pair}, created by the
interaction with the laser field, recombines to produce a photon and
there is no electron transfer from/to the molecule. Thus CIF involves many-body states of the molecule with different total charges whereas LIF only depends on states with
the same charge.

The system is described by the
Hamiltonian $\hat{H}=\hat{H}_{m}+\hat{H}_{f}+\hat{H}_{A}+\hat{H}_{B}+\hat{H}_{mf}+\hat{H}^{\prime }$
where $\hat{H}_{m}$, $\hat{H}_{f}$, $\hat{H}_{A}$ and $\hat{H}_{B}$
represent the independent molecule, radiation field, lead $A$ and lead $B$, respectively.
$\hat{H}_{mf}$ is the coupling between the radiation field and the molecule
\begin{equation}
\label{m-f-coupling}
\hat{H}_{mf}=\hat{a}_{s}^{\dag }\hat{B}+\hat{B}^{\dag }\hat{a}_{s}.
\end{equation}
 $\hat{a}_{s}^{\dag }(\hat{a}_{s})$ are, respectively, the creation
(annihilation) operators for the $s$-th mode of the scattered field
and $\hat{B}^{\dag }(\hat{B})$ is an exciton (electron-hole pair) operator for
the molecule
\begin{eqnarray}
\hat{B}^\dag = \sum_{i>j} \mu_{ij}\hat{\psi}^\dag_i\hat{\psi}_j,~~~
\hat{B} = \sum_{i>j} \mu_{ji}\hat{\psi}^\dag_j\hat{\psi}_i.
\end{eqnarray}
$\mu_{ij}$ is the dipole matrix element between the orbital $i$ and the orbital $j$.

The molecule-lead coupling $H^{\prime}$ is
\begin{equation}
\label{h-prime}
\hat{H}^{\prime }=\sum_{x\in A,B}\sum_{i}J_{xi}\hat{\psi} _{i}^{\dag }\hat{\psi}_{x}+h.c.
\end{equation}

The signal is defined by the expectation value of the flux operator
for the scattered photon modes,
\begin{equation}
\label{sig}
S(\omega _{s},t)=\langle \frac{d}{dt}\hat{a}_{s}^{\dag }\hat{a}_{s}\rangle =i\langle
\lbrack \hat{H},\hat{a}_{s}^{\dag }\hat{a}_{s}]\rangle.
\end{equation}
The angular brackets denote the trace with respect to the
full density matrix of the field+molecule+leads system

Expanding Eq. (\ref{sig}) perturbatively in the emitted field and in the
lead-molecule coupling, the CIF signal from a negatively charged molecule
can be expressed in terms of Liouville space Green's functions. Depending on the
time ordering of various superoperators corresponding to $\hat{B}$, $\hat{a}$ and
$\hat{\psi}$, the density matrix evolves through different LSPs.
\begin{figure}[h]
\centering
\rotatebox{0}{\scalebox{.8}{\includegraphics{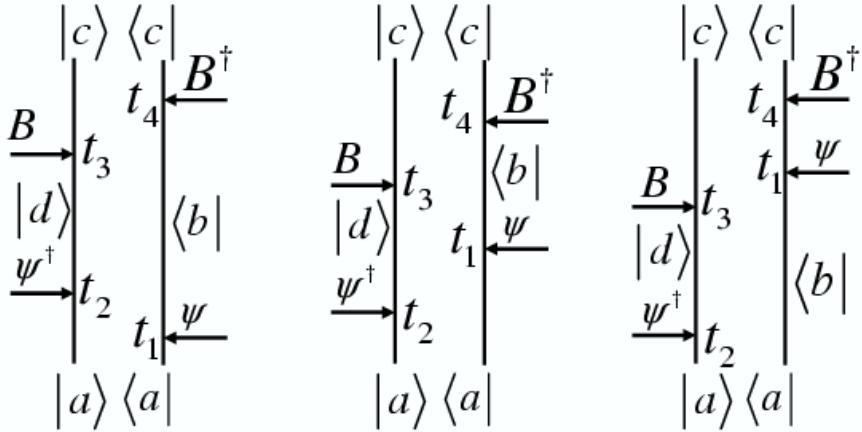}}}\newline
\caption{ Three Liouville space pathways that contribute to the fluorescence from a negatively charged
molecule, Eq. (\ref{fluorescence}). States $|a\rangle$, $|b\rangle$, $|c\rangle$ and $|d\rangle$
denote different charge states of the molecule. Time increases from bottom to top.
}
\label{fig3}
\end{figure}
For the signal to be finite, we must first prepare the molecule in one of its excited states
by transferring either an electron or a hole, before bringing it down to lower excited states.
This results in the cancelation of many terms in the perturbative expansion
of Eq. (\ref{sig}). Only three LSPs (plus their complex conjugates) contribute to the signal.
These can be expressed by the double
sided Feynman diagrams shown in Fig. 3. The fluorescence from a
negatively charged molecule is finally given by \cite{up-jeremy-prb2}
\begin{eqnarray}
\label{fluorescence}
&&\frac{dS_{n}}{dV}(\omega _{s})
=ie\sum_{i<j}\sum_{i^{\prime }<j^{\prime }}\sum_{kl}\sigma
(E_{F}+eV)J_{k}(E_{F}+eV)J_{l}(E_{F}+eV)\nonumber\\
&\times& \mu _{ij}\mu _{i^{\prime
}j^{\prime }}\int d\tau \int_{-\infty }^{t-\tau }d\tau
_{1}\int_{-\infty }^{t}d\tau _{2}
e^{-i\omega _{s}\tau }e^{-i(E_{F}+eV)(\tau _{1}-\tau _{2})}\nonumber\\
&&\left[ \frac{{}
}{{}}G_{LL}^{ji}(t-\tau ,t-\tau ^{+})\left( G_{RR}^{{i^{\prime }}{j^{\prime }%
}}(t,t^{+})G_{RL}^{lk}(\tau _{2},\tau _{1})-G_{RL}^{l{j^{\prime }}}(\tau
_{2},t)G_{RL}^{{i^{\prime }}k}(t,\tau _{1})\right) \right.   \nonumber \\
&-&\left. G_{RL}^{{i^{\prime }}i}(t,t-\tau )\left( G_{RL}^{jk}(t-\tau ,\tau
_{1})G_{RL}^{l{j^{\prime }}}(\tau _{2},t)+G_{LR}^{j{j^{\prime }}}(t-\tau
,t)G_{RL}^{lk}(\tau _{2},\tau _{1})\right) \right.   \nonumber \\
&+&\left.G_{RL}^{li}(\tau _{2},t-\tau )\left( G_{LR}^{j{j^{\prime }}}(t-\tau
,t)G_{RL}^{{i^{\prime }}k}(t,\tau _{1})+G_{RL}^{jk}(t-\tau ,\tau
_{1})G_{RR}^{{i^{\prime }}{j^{\prime }}}(t,t^{+})\right)\right]
\end{eqnarray}
where $\sigma(E)=-\mbox{Im}G_{--}(E)/\pi$ is the density of states of the lead from which electron has been
transferred to the molecule, evaluated at energy $E_F+eV$. $J_k(E)$ is the
coupling of the $k$th molecular orbital with the state of the lead at energy
$E$. A similar expression can be obtained for the signal from a positively
charged molecule\cite{up-jeremy-prb2}.


\subsection{ Electroluminescence: Fluorescence induced by electron and hole transfer}

Electroluminescence is the recombination of an electron and a hole, injected separately by external charge sources.
This is a higher order process than described by Eq. (\ref{fluorescence}), as both electron and hole are injected into the molecule from the leads.
In a molecule-lead configuration, as
discussed in Eqs. (\ref{m-f-coupling}) and (\ref{h-prime}), a lowest
order perturbative expression can be obtained by expanding Eq.
(\ref{sig}) to fourth order in the left and/or right lead couplings.
This contains a large number of terms. The process is simpler in the high bias limit ($V>k_BT$)
where an electron is transferred from the left lead to the molecule while a hole is
transferred from the right lead simultaneously and the reverse
processes are not allowed. We then get
\begin{eqnarray}
\label{electrolum}
S_{EL}(\omega_s,t)&=&-8\mbox{Re}\int_{-\infty}^t
d\tau \int d\tau_1\int d\tau_2\int d\tau_3\int d\tau_4 \sum_{ab}
f_a(1-f_b)e^{-i\epsilon_a(\tau_1-\tau_2)}\nonumber\\
&\times&e^{i\epsilon_b(\tau_3-\tau_4)}
e^{-i\omega_s(t-\tau)}
\sum_{ijkl}J_{ia}J_{lb}J_{a j}J_{b k}\nonumber\\
&\times&\langle\breve{T}\breve{B}_L(\tau)\breve{B}_R^\dag(t)
\breve{\psi}_{iL}^\dag(\tau_1)\breve{\psi}_{jR}(\tau_2)
\breve{\psi}_{kL}(\tau_3)\breve{\psi}_{lR}^\dag\rangle
\end{eqnarray}
where the trace is with respect to the density matrix of the
free molecule. The correlation function
inside the integral can be computed using the many-body states of the molecule.
For a free-electron model it can be expressed as sum of the products of the
Green's functions for the molecule using Wick's theorem.


\section{Dyson Equations for the retarded, advanced and correlation Green's
functions}\label{sec-dyson}

In the previous section we demonstrated how various observables of the coupled
system can be expressed in terms of the Green's functions of the individual sub-systems, $G_{\alpha\beta}$.
Here we show how to compute these Green's functions with
many-body (e.g. electron-electron and electron-phonon) interactions.
Using the superoperator approach, a closed equation of motion (EOM) for the
Green's functions is derived directly in PTSA representation.
The retarded Green's function satisfies an independent equation (decoupled
from the other Green's functions). Nonlinear effects due to
interactions with the environment are included in the self-energy
which can be computed perturbatively in the interaction picture or
using the generating function technique \cite{negele,up-shaul-JCP}.

The Hamiltonian of each closed system ($A$ or $B$) will be
partitioned as
\begin{equation}
\hat{H}=\hat{H}_{0}+\hat{H}_{int}
\end{equation}
where
\begin{equation}
\hat{H}_{0}=\sum_{x}\epsilon _{x}{\hat{\psi}^{\dagger
}}_{x}{\hat{\psi}}_{x}
\end{equation}
is the free-electron part where $x$ denotes the
electronic degrees of freedom, e.g., orbital, spin, wavevector,etc.
$\hat{H}_{int}$ represents the
many-body interactions in the system. We need not specify the form of $H_{int}$
at this point.

In terms of the superoperator PTSA representation, the total Liouville operator is
\begin{eqnarray}
\label{liouvilian} L=L_0+L_{int}
\end{eqnarray}
where $L_{0}$ corresponds to $\hat{H}_0$
\begin{eqnarray}
\label{L0}
L_0 = \sum_{x}\epsilon_x \left({\breve{\psi}^\dagger}_{x+}{\breve{\psi}}_{x+}-{%
\breve{\psi}}_{x-}{\breve{\psi}^\dagger}_{x-}\right)
\end{eqnarray}
and
\begin{eqnarray}  \label{Lint}
L_{int}&=& \sqrt{2}\breve{H}_{int-}.
\end{eqnarray}

We shall construct the equation of motion (EOM) for the
superoperator, $\breve{\psi}_{i\alpha }$ starting with the superoperator
Heisenberg equation
\begin{equation}
\label{nnnn-1}
\frac{\partial }{\partial t}{\breve{\psi}}_{i\alpha
}=i[L,{\breve{\psi}} _{x\alpha }]
\end{equation}
where $[L,{\breve{\psi}} _{x\alpha }]= L{\breve{\psi}} _{x\alpha }-{\breve{\psi}} _{x\alpha }L$.
Substituting Eq. (\ref{liouvilian}) in (\ref{nnnn-1}) and using
the commutation rules Eq. (\ref{eq-supercommu}) of Fermi
superoperators, we get
\begin{equation}
\label{EOM-1} \frac{\partial }{\partial t}{\breve{\psi}}_{x\alpha
}=-i\epsilon _{x}{\breve{\psi}}_{x\alpha}
-i\sqrt{2}[\breve{H}_{-},{\breve{\psi}}_{x\alpha }].
\end{equation}

Differentiating Eq. (\ref{liouville-green}) and using (\ref{EOM-1}),
we obtain the EOM for the Green's functions,
\begin{eqnarray}
\label{EOM-2}
\left( i\frac{\partial }{\partial t}-
\epsilon _{x}\frac{{}}{{}}\right) G^{x\xp}_{\alpha \beta
}(t,{t^{\prime }})=\delta_{x\xp}\delta_{\alpha \beta }\delta (t-{t^{\prime }})
-i\sqrt{2}\langle \breve{T}[\breve{H}_{-}(t),{\breve{\psi}}_{x\alpha
}(t)]{\breve{\psi}^{\dagger }}_{\xp\beta }({t^{\prime }})\rangle .
\end{eqnarray}

When the many-body interaction $H_{int}$ is turned off, Eq. (\ref
{EOM-2}) reduces to
\begin{equation}
\label{EOM-3} \left( i\frac{\partial }{\partial
t}-\epsilon _{x}\right) G^{0x\xp}_{\alpha \beta }(t,{t^{\prime
}})=\delta _{x\xp}\delta _{\alpha \beta }\delta (t-{t^{\prime }}).
\end{equation}
Equation (\ref{EOM-3}) defines the noninteracting Green's functions,
$G_{\alpha \beta }^{0}$.

Equation (\ref{EOM-2}) can be recast in the Dyson form (repeated
indices are summed over).
\begin{equation}
\label{dyson}
\mathbf{G}(t,{t^{\prime }})=\mathbf{G}^{0}(t,{t^{\prime }})+\mathbf{G}^{0}(t,%
{t^{\prime }}_{1})\mathbf{\Sigma }({t^{\prime }}_{1},{t^{\prime }}_{2})%
\mathbf{G}({t^{\prime }}_{2},{t^{\prime }})
\end{equation}
where $\mathbf{G}$ is the $2\times 2$ matrix of Greens functions,
$G^{\alpha \beta }$ and the self-energy matrix $\mathbf{\Sigma }$ is
defined by the equation
\begin{eqnarray}
&&\Sigma ^{x\xp_1}_{\alpha {\alpha ^{\prime }}}(t,{t^{\prime
}} _{1})G^{\xp_1\xp}_{{\alpha ^{\prime }}\beta }({t^{\prime
}}_{1},{t^{\prime }})
=-i\sqrt{2}\langle \breve{T}[\breve{H}_{-}(t),{\breve{\psi}}_{x\alpha
}(t)]{\breve{\psi}^{\dagger }}_{\xp\beta }({t^{\prime }})\rangle   \nonumber \\
&=&-i\sqrt{2}\left\langle \!\breve{T}[{\breve{H}}^I_{-}(t),{{\breve{
\psi}^I}}_{x\alpha }(t)]{{\breve{\psi}}}_{\xp\beta }^{I\dag }({t^{\prime
}}) \mbox{exp}\left\{ -i\sqrt{2}\int_{\infty }^{t}d\tau
{\breve{H}}^I_{-}(\tau )\right\}\right\rangle _{0}
\label{interaction-self-energy}\nonumber\\
\end{eqnarray}
$\langle \cdots \rangle _{0}$ is the trace with respect to the
non-interacting density matrix of the sub-system and the time
dependence is in the interaction picture with respect to $\hat{H}_0$.

The Dyson equation [Eq. (\ref{dyson})] can be written explicitly in the matrix
form with
\begin{eqnarray}
\label{dyson-matrix}
{\bf G}=\left( \begin{array}{cc}
G_{--}&G_{-+}\\
0&G_{++}
\end{array} \right),~
{\bf G}^0=\left( \begin{array}{cc}
G_{--}^0&G_{-+}^0\\
0&G_{++}^0
\end{array} \right),~
{\bf \Sigma}=
\left( \begin{array}{cc}
\Sigma_{--}&\Sigma_{-+}\\
\Sigma_{+-}&\Sigma_{++}
\end{array} \right),
\end{eqnarray}
where we have used the fact that $G^{+-}=0$, Eq. (\ref{g+-1}).

The bottom left matrix element in Eq. (\ref{dyson-matrix}) gives
\begin{equation}
\label{eq-g4} G^{0}_{++}\Sigma_{+-}G_{--}=0.
\end{equation}
Since $G_{++}$ and $G_{--}$ do not generally vanish, Eq.
(\ref{eq-g4}) implies that $\Sigma_{+-}=0$. Equation
(\ref{dyson-matrix}) then gives the retarded $G_{--}$ Green's
functions thus satisfy their own Dyson equations.
\begin{eqnarray}
G_{--} &=&G^{0}_{--}+G^{0}_{--}\Sigma_{--}G_{--}. \label{eq-ng1}
\end{eqnarray}
Similarly we get for the advanced Green's function
\begin{eqnarray}
G_{++} =G^{0}_{++}+G^{0}_{++}\Sigma_{++}G_{++}  \label{eq-ng2}.
\end{eqnarray}
The correlation function $G^{-+}$ can then be expressed in terms
of the retarded and the advanced Green's functions and their
self-energies
\begin{eqnarray}
G_{-+} &=&G^{0}_{-+}+G^{0}_{-+}\Sigma_{++}G_{++}
+G^{0}_{--}\left( \Sigma ^{--}G_{-+}+\Sigma_{-+}G_{++}\right) .
\label{eq-ng3}
\end{eqnarray}

Keldysh had derived Eqs. (\ref{eq-ng1})- (\ref{eq-ng3})
\cite{keldysh,haug-jauho} by starting with the coupled equations for
the four Hilbert space Green's functions in the PTBK representation
(which in superoperator notation correspond to $G_{LL}, G_{RR}, G_{RL}$ and $G_{LR}$)
and then decouple them by transforming to the PTSA representation.
The superoperator
approach allows us to work with the PTSA representation from the outset.
The derivation is much more compact and physically-transparent.

For completeness, we present the corresponding equations for bosons in Appendix \ref{boson}.

\section{Current-fluctuation statistics}\label{statistics}

So far we have discussed average observables, such as the total energy,
change in the charge density and the current in interacting
open systems. Fluctuations may be measured in
electron tunneling between two conductors and electron transport through
quantum junctions \cite{Rammer}. Much theoretical\cite{statistics-theory} and
experimental\cite{statistics-experiment} effort has been devoted
recently to the statistical behavior of the
transport through such systems. Here we do not review these works but
rather give a flavor of it. More detailed derivations
are given in Ref. \cite{nazarov,two-point}.

\subsection{Electron-tunneling statistics}

We again consider the two interacting systems $A$ and $B$ described by the Hamiltonian,
Eqs. (\ref{close-1}) and (\ref{hab}).
The number $k$ of electrons transferred in time $t$
can be computed by
measuring the number of electrons in one of the systems at time
$t=0$ and again at time $t$. The difference of
the two measurements, $k$, is a fluctuating variable which changes
as the measurement is repeated over the same time interval.
Thus $k$ has a distribution
$P(k,t)$ whose first moment gives the average
current, Eq. (\ref{eq-final-2-y}). For small applied external bias, the charge
transfer can reverse its course and
these fluctuations may become non-negligible.

We shall compute the statistics by using the generating
function associated with $P(k,t)$
\begin{equation}
\label{gf-definition} \chi (\lambda
,t)\equiv\sum_{k}\mbox{e}^{i\lambda k}P(k,t).
\end{equation}

The GF is a continuous function of the parameter $\lambda $, which is
conjugate to the net number of electron transfers $k$. It is convenient for
computing moments of various orders by
taking derivatives of GF with respect to $\lambda $. Working directly with
$P(k,t)$  requires summations over the discrete
variable $k$ from $-\infty $ to $\infty $.
$P(k,t)$ is obtained by the inverse Fourier transform of the
generating function
\begin{eqnarray}
\label{prob-four}
P(k,t)=\int_0^{2\pi}\frac{d\lambda}{2\pi}e^{-i\lambda
k}\chi(\lambda,t).
\end{eqnarray}

The $n$-th moment of $P(k,t)$ is given by
\begin{eqnarray}
\langle k^n(t) \rangle = (-i)^n\left.\frac{\partial}{\partial\lambda^n}\chi(\lambda,t)\right|_{\lambda=0}.
\end{eqnarray}
The $n$'th order cumulants are certain combinations of the moments of order $n$ and lower.
The first cumulant ${\cal C}^{(1)}=\langle k(t)\rangle$ is the first moment and
gives the average number of particles transferred from $A$ to $B$.
The second cumulant ${\cal C}^{(2)}=\langle (k(t)-\langle k(t)\rangle)^2\rangle$ is
the simplest measure of fluctuations of $k$ around its average value. The third cumulant
${\cal C}^{(3)}=\langle (k(t)-\langle k(t)\rangle)^3\rangle$ gives the
skewness (asymmetry) of $P(k,t)$. For a Gaussian distribution
all cumulates higher than the first two vanish. The higher cumulants
thus measure the non Gaussian nature of the distribution.
All cumulants of the current statistics can be computed directly by
taking the derivatives of the cumulant generating function,
$Z(\lambda,t)=\mbox{log}\chi(\lambda,t)$, with respect to $\lambda $.
The $n$-th cumulant is given by
\begin{equation}
\label{cumulant}
\mathcal{C}^{(n)}(t)=(-i)^n\left. \frac{\partial ^{n}}{\partial \lambda^n }%
Z(\lambda,t)\right\vert _{\lambda =0}.
\end{equation}


For our model, the generating function, Eq.
(\ref{gf-definition}), can be expressed as\cite{two-point}
\begin{equation}
\chi (\lambda ,t)=\int_{0}^{2\pi }\frac{d\Lambda }{2\pi }\mbox{e}^{Z(\lambda
,\Lambda ;t)}
\end{equation}
with
\begin{equation}
\label{cum-gf}
Z(\lambda ,\Lambda ;t)=\mbox{log}\left[ \mbox{Tr}\left\{ \mathcal{G}_{I}(\lambda,\Lambda,t)\rho_0\right\} \right]
\end{equation}
and $\mathcal{G}_{I}$ is the interaction picture propagator given in
Appendix \ref{mod-evol}.

We shall evaluate Eq. (\ref{cum-gf}) perturbatively to second order in the coupling,
Eq. (\ref{hab}). As was done in Sec. \ref{gf-nn}, in the infinite past, the density matrix of the two systems is factorized,
$\rho_0=\rho_A\oplus\rho_B$.  The interactions are then built-in by adiabatic switching.
We then express the cumulant GF, Eq. (\ref{cum-gf}), in terms of the
Green's functions of systems $A$ and $B$ (see Appendix
\ref{mod-evol}).
\begin{eqnarray}
\label{gf-final}
Z(\lambda ,\Lambda ;t)&=& 2i\mbox{sin}\left( \frac{\lambda }{2}
\right)\int_{-\infty }^{t}d\tau _{1}\int_{0}^{t}d\tau
_{2}\left\{ \theta (\tau _{1})\mathcal{F}(\lambda ,\tau _{1},\tau
_{2})\right.   \nonumber  \\
&+&\left. \theta (-\tau _{1})\left[ e^{-i\Lambda }\mathcal{F}(0,\tau
_{1},\tau _{2})+e^{i\Lambda }\mathcal{F}(0,\tau _{2},\tau _{1})\right]
\right\}
\end{eqnarray}
with
\begin{eqnarray}
\label{function-f}
\mathcal{F}(\lambda ,t,{t^{\prime }}) &=&\sum_{a{a^{\prime }}b{{b^{\prime }}}
}J_{ab}J_{{a^{\prime }}{{b^{\prime }}}}^{\ast }\left[ e^{-i\frac{\lambda }{2}
}G_{RL}^{{a^{\prime }}a}(t,{t^{\prime }})G_{LR}^{b,{{b^{\prime }}}}({
t^{\prime }},t)
- e^{i\frac{\lambda }{2}}G_{RL}^{b,{{b^{\prime }}}}(t,{t^{\prime }}
)G_{LR}^{{a^{\prime }}a}({t^{\prime }},t)\right] .
\end{eqnarray}
All Green's functions appearing in Eqs. (\ref{int-energy-j2}),
(\ref{charge-density}) and (\ref{function-f}) correspond to the
isolated sub-systems. These
can include many-body interactions $e-e$ or $e-p$. The Green's functions then include
self-energy due to many-body interactions as shown in the previous section.
The self-energy to first order in $e-e$ interactions is given in Appendix \ref{se-ee}.

\subsection{Current fluctuations: electron-transport statistics}

We now consider electron transport between two leads $A$ and
$B$ through a quantum system such as a molecule or a quantum dot.
The total system Hamiltonian is
\begin{eqnarray}
\label{hamil-2} \hat{H}_T= \hat{H}_A+ \hat{H}_B+ \hat{H}_S+ \hat{V}
\end{eqnarray}
where $ \hat{H}_A$ and $ \hat{H}_B$ represent Hamiltonian of two
leads $A$ and $B$, and $ \hat{H}_S$ is the system Hamiltonian.
We assume a free- electron gas model for all sub-systems.
\begin{eqnarray}
\label{eq-111}
 \hat{H}_X=\sum_{x\in X}\epsilon_x\ \hat{\psi}^\dag_x \hat{\psi}_x, ~~~X=A,B,S.
\end{eqnarray}
$ \hat{V}$ represents the coupling Hamiltonian between the molecule and the
leads.
\begin{eqnarray}
\label{eq-112}
 \hat{V}=\sum_{x\in A,B}J_{xi} \hat{\psi}^\dag_{x} \hat{\psi}_i + h.c.
\end{eqnarray}
where $J_{xi}$ is the coupling between the lead ($x$) and the system ($i$)
orbitals.

A simple perturbative calculation in the coupling, as was done in
going from Eq. (\ref{cum-gf}) to (\ref{cum-gf-11}),
will miss the non-equilibrium state of the
embedded system. However, using the path-integral
technique\cite{two-point}, the cumulant GF $Z(\lambda,\Lambda)$ can
be expressed in terms of SNGF of the molecule. The final expression
for the GF involves determinant of the matrix Green's functions
renormalized due to the coupling with $A$ and $B$. For long counting
times, the cumulant GF is given by \cite{two-point}
\begin{eqnarray}
Z(\lambda,\Lambda) = \int d\omega~ \mbox{ln} \mbox{Det}[{\bf
G}^{-1}(\omega,\lambda)]
\end{eqnarray}
where the $2\times 2$ matrix ${\bf G}(\omega,\lambda)$ is obtained by solving a Dyson
matrix like equation [Eq. (\ref{dyson-matrix})] with the
following self-energy matrix
\begin{eqnarray}
\Sigma^{ii^\prime}_{++}(\omega,\lambda)&=& \frac{i}{2}\Gamma^B_{ii^\prime}
+\frac{i}{2}\Gamma^A_{ii^\prime}\left[e^{i\lambda}(1-f_A(\omega))
+ e^{-i\lambda}f_A(\omega)
\right]\label{sig++}\\
\Sigma^{ii^\prime}_{--}(\omega,\lambda)&=& -\frac{i}{2}\Gamma^B_{ii^\prime}
-\frac{i}{2}\Gamma^A_{ii^\prime}\left[e^{i\lambda}(1-f_a(\omega))
+e^{-i\lambda}f_A(\omega)
\right]\label{sig--}\\
\Sigma^{ii^\prime}_{+-}(\omega,\lambda)&=&
-\frac{i}{2}\Gamma^A_{ii^\prime}\left[(e^{i\lambda}-1)(1-f_A(\omega)
-(e^{-i\lambda}-1)f_A(\omega))
\right]\label{sig+-}\\
\Sigma^{ii^\prime}_{-+}(\omega,\lambda)&=&
-i\Gamma_{ii^\prime}^B(2f_B(\omega)-1)+\!\!\frac{i}{2}\Gamma_{ii^\prime}^A\nonumber\\
&&\left[(e^{i\lambda}+1)(1-f_A(\omega))-(e^{-i\lambda}+1)f_A(\omega)\right]\label{sig-+}
\end{eqnarray}
here $\Gamma_{ii^\prime}^X=2\pi\sum_{x\in X}J_{xi}J_{i^\prime x}
\delta(\omega-\epsilon_x)$ comes from interactions with the leads.
When $\lambda=0$, $\Sigma_{+-}=0$ as
discussed previously (Eq. (\ref{eq-g4})), and $\Sigma_{--}$, $\Sigma_{++}$ and
$\Sigma_{-+}$ reduce to ordinary retarded, advanced and correlation
self-energies due to leads-molecule interaction, respectively.


\section{Conclusions}\label{conclusion}

We have presented a superoperator Liouville space nonequilibrium Green's function (SNGF) theory in the
PTSA ($+/-$) representation for interacting open many-electron systems.

By constructing the microscopic equations of motion for superoperators in PTSA representation,
the response and correlation functions are
treated along the same footing in Liouville space. This is reminiscent of
the MSR formulation\cite{msr} of classical statistical mechanics.
The various Green's functions of the CTPL formulation
appear naturally as a consequence of a simple
superoperator time ordering operation which controls the evolution of the bra and
the ket of the density matrix in Liouville space. The interaction energy
and the change in density matrix of individual sub-systems are
directly recast in terms of the retarded, advanced and correlation
Green's functions. Electron-transfer statistics in both
tunneling and transport may be formulated in
terms of the SNGF. Many-body effects (such as $e-e$ interactions)
can be included perturbatively through self-energies by solving the Dyson
equations.
A notable advantage of the PTSA representation is that since $G_{+-}=0$,
the matrix Dyson equations of the other three Green's functions are decoupled.
Dyson equations for the retarded the and advanced Green's functions are derived
without computing first the matrix Dyson equation for four Keldysh
Green's functions in PTBK representation. This is because the linear transformation which
results in PTSA operators can be performed at the operator level rather
than on the Green's functions or the generating function.
In Liouville space we can work directly with the
operators in PTSA representation. In Hilbert space (or PTBK), in contrast,
one has to first compute the four (lesser, greater,time ordered,
anti-time ordered) Green's functions and only then obtain the
retarded and advanced Greens functions using the matrix
transformation. This is not necessary in Liouville space.
The superoperator Green's function formulation for bosons is given in
Appendix \ref{boson} for completeness.

\section*{Acknowledgments}

The support of the National Science Foundation (Grant No. CHE-0745892)
and NIRT (Grant No. EEC 0303389)) is gratefully acknowledged.

\appendix

\section{SNGF for bosons}\label{boson}

We denote the boson creation and annihilation
operators by $\hat{\phi}_i$ and $\hat{\phi}_i^\dag$, respectively.
These satisfy the commutation relations
\begin{eqnarray}
\label{boson-commutation}
[\hat{\phi}_i, \hat{\phi}_j^\dag] &=& \delta_{ij},~~~
[\hat{\phi}_i, \hat{\phi}_j]= [\hat{\phi}_i^\dag, \hat{\phi}_j^\dag] = 0
\end{eqnarray}
the corresponding superoperators, $\breve{\phi}_\alpha$ and
$\breve{\phi}_\alpha^\dag$, in Liouville space are defined by Eq. (\ref{liouv-operator})
\begin{eqnarray}
\label{boson-definition}
\breve{\phi}_L X &\equiv& \hat{\phi}X,~~
\breve{\phi}_L^\dag X \equiv \hat{\phi} X
\nonumber \\
\breve{\phi}_R X &\equiv& X \hat{\phi},~~
\breve{\phi}_R^\dag X \equiv X\hat{\phi}^\dag.
\end{eqnarray}
A notable difference between Eqs. (\ref{boson-definition}) and (\ref{eq-3aa}) is
the absence of the $(-1)$ factors for bosons.
Boson superoperators in PTSA representation ("+" and "-") are defined by
Eqs. (\ref{trans}) and satisfy commutation relations similar to Eqs.
(\ref{boson-commutation}).
\begin{eqnarray}
\label{boson-comm-2}
[\breve{\phi}_{m\alpha},\breve{\phi}_{n\beta}^\dag]=\delta_{mn}\delta_{\alpha\beta},~
[\breve{\phi}_{m\alpha},\breve{\phi}_{n\beta}]=
[\breve{\phi}^\dag_{m\alpha},\breve{\phi}^\dag_{n\beta}]=0.
\end{eqnarray}
Using Eqs. (\ref{boson-comm-2}) in  (\ref{pm-1}) and (\ref{pm}), we
get
\begin{eqnarray}
(\hat{\phi}_i\hat{\phi}_j^\dag)_-&=&
\frac{1}{\sqrt{2}}\left[\delta_{ij}+\breve{\phi}_{i+}\breve{\phi}_{j-}^\dag+
\breve{\phi}_{i-}\breve{\phi}_{j+}^\dag\right]\label{11}\\
(\hat{\phi}_i\hat{\phi}_j^\dag)_+&=&
\frac{1}{\sqrt{2}}\left[\delta_{ij}+\breve{\phi}_{i+}\breve{\phi}_{j+}^\dag+
\breve{\phi}_{i-}\breve{\phi}_{j-}^\dag\right]\label{22}.
\end{eqnarray}
The corresponding expressions for $(\hat{\phi}_i\hat{\phi}_j)_-$
and $(\hat{\phi}_i\hat{\phi}_j)_+$ can be obtained from Eq.
(\ref{11}) and (\ref{22}) by simply neglecting the $\delta_{ij}$ terms
inside the brackets.

In analogy to Eq. (\ref{liouville-green}), the boson superoperator Green's functions are defined as,
\begin{eqnarray}  \label{boson-green}
D^{mn}_{\alpha\beta}(t,{t^\prime})=-i\langle
\breve{T}\breve{\phi}_{m\alpha}(t)\breve{\phi}_{n\beta}^\dag({t^\prime})\rangle.
\end{eqnarray}

Using Eqs. (\ref{boson-definition}) and (\ref{boson-green}), it is
easy to show that $D_{+-}$, $D_{++}$ and $D_{-+}$ represent the
retarded, advanced and correlation Green's functions, and
$D_{--}=0$. This gives the boson analogue of Eqs. (\ref{g--1})-(\ref{g-+1}).
\begin{eqnarray}  \label{boson-eq}
D_{+-}^{ij}(t,{t^\prime})&=& -i\theta(t-{t^\prime})\langle[%
\breve{\phi}_i(t),\breve{\phi}^\dag_j({t^\prime})]\rangle \\
D_{-+}^{ij}(t,{t^\prime})&=& i\theta({t^\prime}-t)\langle[%
\breve{\phi}_i(t),\breve{\phi}^\dag_j({t^\prime})]\rangle \\
D_{++}^{ij}(t,{t^\prime})&=& -i\langle\{\breve{\phi}_i(t),\breve{\phi}^\dag_j({%
t^\prime})\}\rangle \\
D_{--}^{ij}(t,{t^\prime})&=& 0.
\end{eqnarray}
Unlike Fermions, in this case $D_{--}=0$ while $D_{+-}$
represents the retarded Green's function. This comes from
the different commutation relations, Eq.
(\ref{boson-comm-2}) and (\ref{eq-1a}).
The different particle-statistics for bosons affects the form of the Dyson equation
(Eq. (\ref{dyson-matrix})).

The Dyson equation for bosons, ${\bf D}=\bf{D}^0+\bf{D}^0 {\bf \Xi}~ {\bf D}$, can be derived following the same steps which
lead to  Eq. (\ref{dyson-matrix}). Here
\begin{eqnarray}
\label{dyson-boson} 
{\bf D}=\left( \begin{array}{cc}
0&D_{-+}\\
D_{+-}&D_{++}
\end{array} \right),~
{\bf D}^0=\left( \begin{array}{cc}
0&D_{-+}^0\\
D_{+-}^0&D_{++}^0
\end{array} \right),~
{\bf \Xi}=\left( \begin{array}{cc}
\Xi_{--}&\Xi_{-+}\\
\Xi_{+-}&\Xi_{++}
\end{array} \right),
\end{eqnarray}
where $D^{0}$ is the Green's function for the reference
system and $\Xi$ is the self-energy due to many-body interactions.

Comparing the top right matrix elements on both sides then gives,
$\Xi_{++}=0$. We finally get the boson analogues of Eqs. (\ref{eq-g4})-(\ref{eq-ng3}).
\begin{eqnarray}
D_{+-}&=&D^0_{+-}+D_0^{+-}\Xi_{-+}D_{+-} \label{bgf-1}\\
D_{-+}&=&D^0_{-+}+D_0^{-+}\Xi_{+-}D_{-+} \label{bgf-2}\\
D_{++}&=&D^0_{++}\Xi_{+-}D_{-+}
+D_{+-}[\Xi_{--}D_{-+}+\Xi_{-+}D_{++}]\label{bgf-3}.
\end{eqnarray}
These equations are decoupled and each of the Green's functions can be computed separately. $D_{+-}$ and $D_{-+}$ represent the retarded and the advanced boson Green's functions.

\section{Wick's Theorem}\label{wick}

Wick's theorem states that the expectation value of the product of
$n$ fermion field operators with respect to a many-body state, represented by a single Slater determinant, can be expressed as the sum of the expectation values
of products of all possible pairs of
operators \cite{mahan-book,mills,thauless,negele}. This theorem provides a practical
tool for factorizing the expectation value of a time ordered product of
operators to simpler quantities. Note that expectation value vanishes unless $n$ is even.

This theorem can be derived as follows.
Since a single determinant represents the eigenstate of
corresponds to a noninteracting Hamiltonian, $H_{0}=\sum_{ij}K_{ij}\hat{A}%
_{i}\hat{A}_{j}$, the expectation of a product of $n$ operators
$\langle \hat{A}_{1}\hat{A}_{2}\cdots \hat{A}_{n}\rangle $ involves
an integral
weighted by a Gaussian operator $\sim\mbox{exp}\left( -\beta \sum_{ij}
\hat{A}_{i}K_{ij}\hat{A}_{j}\right) $. Using properties of Gaussian
integrals, it can be factorized as \cite{Zinn-Justin,mills}.
\begin{equation}
\label{expectation}
 \langle \hat{A}_{1}\hat{A}_{2}\cdots
\hat{A}_{n}\rangle =\sum_{P}\langle \hat{A}_{1}\hat{A}_{2}\rangle
\cdots \langle \hat{A}_{n-1}\hat{A}_{n}\rangle
\end{equation}
where the sum over $P$ runs over all possible pairings of
$\{1,2,\cdots ,n\}$, Equation (\ref{expectation}) remains valid also
for superoperators. By simply replacing
all $\hat{A}_{n^{\prime}}s$ by $\breve{A}_{n\alpha },\alpha =L,R,+,-$.
\begin{eqnarray}
\label{expectation1}
\langle \breve{A}_{1\alpha}\breve{A}_{2\beta}\cdots \breve{A}_{n\gamma}\rangle =
\sum_{P}\langle \breve{A}_{1\alpha}\breve{A}_{2\beta}\rangle
\cdots \langle \breve{A}_{n-1\delta}\breve{A}_{n\gamma}\rangle.
\end{eqnarray}
Since each permutation of fermion superoperators brings a minus sign, the Wick's
theorem for fermion superoperators takes the form,
\begin{eqnarray}
\label{expectation2}
\langle \breve{A}_{1\alpha}\breve{A}_{2\beta}\cdots \breve{A}_{n\gamma}\rangle =
\sum_{P}(-1)^p\langle \breve{A}_{1\alpha}\breve{A}_{2\beta}\rangle
\cdots \langle \breve{A}_{n-1\delta}\breve{A}_{n\gamma}\rangle
\end{eqnarray}
where $p$ is the number of permutations required to get the right pairing.
The same theorem applies to bosons by simply eliminating the $(-1)^p$ factors.

\section{Connecting the closed-time loop with the real time Green's functions}\label{H-L-GF}

Standard CTPL Green's function theory is formulated in terms of four Hilbert space
Green functions: time ordered $(G^{T})$, anti-time ordered
$(G^{\tilde T})$, greater ($G^{>}$) and lesser
$(G^{<})$\cite{keldysh,haug-jauho,mahan-book}. These are defined in the
Heisenberg picture as,
\begin{eqnarray}
\label{hilbert-green}
G^T(t_1,t_2)&\equiv&-i\langle T \hat{\psi}(t_1) \hat{\psi}%
^\dagger(t_2) \rangle  \nonumber \\
&=& -i\theta(t_1-t_2)\langle \hat{\psi}(t_1) \hat{\psi}%
^\dagger(t_2) \rangle
+ \theta(t_2-t_1) \langle \hat{\psi}^\dagger(t_2) \hat{\psi}(t_1)
\rangle
\nonumber \\
G^{\tilde T}(t_1,t_2)&\equiv&-i\langle \tilde{T} \hat{\psi}%
(t_1) \hat{\psi}^\dagger(t_2) \rangle  \nonumber \\
&=& -i\theta(t_2 -t_1)\langle \psi(t_1) \hat{\psi}%
^\dagger(t_2) \rangle
+ \theta(t_1- t_2) \langle \hat{\psi}^\dagger(t_2) \hat{\psi}(t_1)
\rangle
\nonumber \\
G^>(t_1,t_2) &\equiv&-i\langle \hat{\psi}(t_1) \hat{\psi}%
^\dagger(t_2) \rangle  \nonumber \\
G^<(t_1,t_2) &\equiv& i\langle \hat{\psi}^\dagger(t_2) \hat{\psi%
}(t_1) \rangle
\end{eqnarray}
$T$ ($\tilde T$) is the Hilbert space time (anti-time) ordering
operator,  Eqs. (\ref{aaa-1}) and (\ref{aaa-2}).

The four Liouville space Green's functions which appear naturally in
the superoperator formulation are
\begin{eqnarray}
\label{liouville-green-2}
G_{LL}(t_1,t_2)&=& - i\langle\breve{T}
\breve{\psi}_L(t_1) \breve{\psi}^\dagger_L(t_2)\rangle,~
G_{RR}(t_1,t_2)= - i\langle\breve{T}
\hat{\psi}_R(t_1) \breve{\psi}^\dagger_R(t_2)\rangle
\nonumber \\
G_{LR}(t_1,t_2)&=& - i\langle\breve{T}
\breve{\psi}_L(t_1) \breve{\psi}^\dagger_R(t_2)\rangle,~
G_{RL}(t_1,t_2)= - i\langle\breve{T}
\breve{\psi}_R(t_1) \breve{\psi}^\dagger_L(t_2)\rangle
\end{eqnarray}

To establish the connection between Eqs. (\ref{hilbert-green}) and (\ref{liouville-green-2}),
we shall convert the superoperators back to ordinary operators by using
their definitions and  anti-commutation relations. For $G_{LR}$
and $G_{RL}$, we obtain,
\begin{eqnarray}
G_{LR}(t_{1},t_{2}) &\equiv& - i \mbox{Tr}\{\breve{T}
\breve{\psi}_{L}(t_{1}) \breve{\psi}^{\dagger}_{R}(t_{2}) \rho\}  \nonumber \\
&=& - i\left[ \theta(t_{1}-t_{2})\mbox{Tr}\{
\breve{\psi}_{L}(t_{1}) \breve{\psi}^{\dagger}_{R}(t_{2})\rho\}
 - \theta(t_{2}-t_{1})\mbox{Tr}\{\breve{\psi}^{\dagger}_{R}(t_{2})
\breve{\psi}_{L}(t_{1})\rho\}\right]  \nonumber \\
&=& i \mbox{Tr}\{\hat{\psi}(t_{1})
\rho\hat{\psi}^{\dagger}(t_{2})\}=
i \langle\hat{\psi}^{\dagger}(t_{2}) \hat{\psi }%
(t_{1}) \rangle  \nonumber \\
&=& G^{<}(t_1,t_2)  \label{glr} \\
G_{RL}(t_{1},t_{2}) &\equiv& - i \mbox{Tr}\{\breve{T}
\breve{\psi}_{R}(t_{1}) \breve{\psi}^{\dagger}_{L}(t_{2}) \rho\}  \nonumber \\
&=& - i\left[ \theta(t_{1}-t_{2})\mbox{Tr}\{
\breve{\psi}_{R}(t_{1}) \breve{\psi}^{\dagger}_{L}(t_{2})\rho\}
- \theta(t_{2}-t_{1})\mbox{Tr}\{\breve{\psi}^{\dagger}_{L}(t_{2})
\breve{\psi}_{R}(t_{1})\rho\}\right]  \nonumber \\
&=& i \mbox{Tr}\{ \hat{\psi}^{\dagger}(t_{2}) \rho\hat{\psi }
(t_{1}) \} = i \langle\hat{\psi}(t_{1})
\hat{\psi}^{\dagger
}(t_{2}) \rangle  \nonumber \\
&=& - G^{>}(t_{1},t_{2}) \label{grl}
\end{eqnarray}
where $\rho$ is the fully interacting many-body density matrix.

For $G_{LL}$ and $G_{RR}$ we need to distinguish between two
cases,\newline (i). For $t_{1} > t_{2}$, we get,
\begin{eqnarray}
\label{gll&gr1} G_{LL}(t_{1},t_{2}) & \equiv& - i
\mbox{Tr}\{\breve{T} \breve{\psi}_{L}(t_{1}) \breve{\psi}^{\dagger}_{L}(t_{2}) \rho\}  \nonumber \\
& = & - i \mbox{Tr}\{\hat{\psi}(t_{1}) \hat{\psi}%
^{\dagger}(t_{2}) \rho\}
= - i \langle\hat{\psi}(t_{1})
\hat{\psi}^{\dagger}(t_{2})
\rangle  \nonumber \\
G_{RR}(t_{1},t_{2}) & \equiv& - i \mbox{Tr}\{\breve{T}
\breve{\psi}_{R}(t_{1}) \breve{\psi}^{\dagger}_{R}(t_{2}) \rho\}  \nonumber \\
& = &- i \mbox{Tr}\{\rho\hat{\psi}^{\dagger}(t_{2}) \hat{\psi }
(t_{1})\}  = - i \langle\hat{\psi}^{\dagger}(t_{2})
\hat{\psi}(t_{1}) \rangle
\end{eqnarray}
(ii) For the opposite case, $t_{1} < t_{2} $, we get,

\begin{eqnarray}
\label{gll&gr2}
G_{LL}(t_{1},t_{2}) & \equiv & - i \mbox{Tr}\{\breve{T}
\breve{\psi}_{L}(t_{1}) \breve{\psi}^{\dagger}_{L}(t_{2}) \rho\}  \nonumber \\
& = &i \mbox{Tr}\{ \hat{\psi}^{\dagger}(t_{2}) \hat{\psi}
(t_{1}) \rho\}= i \langle\hat{\psi}^{\dagger}(t_{2}) \hat{\psi}
(t_{1}) \rangle  \nonumber \\
G_{RR}(t_{1},t_{2}) & \equiv & - i \mbox{Tr}\{\breve{T}
\breve{\psi}_{R}(t_{1}) \breve{\psi}^{\dagger}_{R}(t_{2})\rho\}  \nonumber \\
& = &i \mbox{Tr}\{\rho\hat{\psi}(t_{1})
\hat{\psi}^{\dagger }(t_{2})\}= i
\langle\hat{\psi}(t_{1}) \hat{\psi}^{\dagger }(t_{2})\rangle
\end{eqnarray}
Combining Eqs. (\ref{gll&gr1}) and (\ref{gll&gr2}) we can write,
\begin{eqnarray}  \label{gll&grr}
G_{LL}(t_{1},t_{2}) & \equiv& - i \mbox{Tr}\{\breve{T}
\breve{\psi}_{L}(t_{1}) \breve{\psi}^{\dagger}_{L}(t_{2}) \rho\}  \nonumber \\
& = &-i \left[ \theta(t_{1}-t_{2})
\langle\hat{\psi}(t_{1})
\hat{\psi}^{\dagger}(t_{2}) \rangle
 - \theta(t_{2}-t_{1}) \langle\hat{\psi}^{\dagger}(t_{2}) \hat{\psi }%
(t_{1}) \rangle\right]  \nonumber \\
&=& G^{T}(t_{1},t_{2})  \nonumber \\
G_{RR}(t_{1},t_{2}) & \equiv & - i \mbox{Tr}\{\breve{T}
\breve{\psi}_{R}(t_{1}) \breve{\psi}^{\dagger}_{R}(t_{2}) \rho\}  \nonumber \\
& = & -i \left[ \theta(t_{1}-t_{2})
\langle\hat{\psi}^{\dagger
}(t_{2}) \hat{\psi}(t_{1}) \rangle
 - \theta(t_{2}-t_{1}) \langle\hat{\psi}(t_{1})
\hat{\psi}^{\dagger
}(t_{2}) \rangle\right]  \nonumber \\
& = & -G^{\tilde T}(t_{1},t_{2})
\end{eqnarray}
Equations (\ref{glr})-(\ref{gll&grr}) show the equivalence of the
Hilbert and the Liouville space Green's functions.


\section{The Retarded, Advanced and Correlation Liouville-space
Green's functions}\label{H-L-GF-2}

Here we show that the SNGFs $G_{--}$,
$G_{++}$ and $G_{-+}$ defined as,
\begin{equation}
G^{ij}_{\alpha\beta}(t,{t^{\prime }})\equiv -i\langle
\breve{T}\breve{\psi}_{i\alpha}(t)\breve{\psi}_{j\beta}^{\dag }({t^{\prime
}})\rangle
\end{equation}
coincide with the retarded, advanced and correlation functions,
respectively.

Starting with $G_{--}$, we expand the time
ordering operator as
\begin{eqnarray}
G^{ij}_{--}(t,{t^{\prime}})
&=& -i\langle\breve{T}\breve{\psi}
_{i-}(t)\breve{\psi}_{j-}^{\dag}({t^{\prime}})\rangle  \nonumber  \label{eq-200} \\
& =& -i\left[
\theta(t-{t^{\prime}})\langle\breve{\psi}_{i-}(t)\breve{\psi}
_{j-}^{\dag}({t^{\prime}})\rangle
- \theta({t^{\prime}}-t)\langle\breve{\psi}_{j-}^{\dag}({t^{\prime}}
)\breve{\psi}_{i-}(t)\rangle\right] .
\end{eqnarray}
Substituting
$\breve{\psi}_{-}=(\breve{\psi}_{L}-\breve{\psi}_{R})/\sqrt{2}$, we
can write
\begin{eqnarray}
\label{eq-3}
\langle\breve{\psi}_{i-}(t)\breve{\psi}_{j-}^{\dag}({t^{\prime}})\rangle
&=& \frac{1}{2}[\langle(\breve{\psi}_{iL}(t)-\breve{\psi}_{iR}(t))(\breve{\psi}_{jL}^{\dag}({%
t^{\prime }})-\breve{\psi}_{jR}^{\dag}({t^{\prime}}))\rangle]  \nonumber \\
&=&\langle\breve{\psi}_{iL}(t)\breve{\psi}_{jL}^{\dag}({t^{\prime}})\rangle+\langle\breve{\psi}
_{iR}(t)\breve{\psi}_{jR}^{\dag}({t^{\prime}})\rangle  \nonumber \\
&-&\langle\breve{\psi}_{iL}(t)\breve{\psi}_{jR}^{\dag}({t^{\prime}})\rangle\!-\!\langle
\breve{\psi}_{iR}(t)\breve{\psi}_{jL}^{\dag}({t^{\prime}})\rangle.
\end{eqnarray}

Using the definitions of $\breve{\psi}_{L}$ and $\breve{\psi}_{R}$ [Eq.
(\ref{eq-3aa})], we get
\begin{eqnarray}
 \label{eq-4}
\langle\breve{\psi}_{i-}(t)\breve{\psi}_{j-}^{\dag}(t^{\prime})\rangle
&\equiv& \langle
\hat{\psi}_{i}(t)\hat{\psi}_{j}^{\dag}({t^{\prime}})\rangle+\langle\hat{\psi}_{j}^{\dag }({%
t^{\prime}})\hat{\psi}_{i}(t)\rangle
= \langle\hat{\psi}_{i}(t),\hat{\psi}_{j}^{\dag}({t^{\prime}})\rangle.
\end{eqnarray}
Similarly,
\begin{eqnarray}
\langle\breve{\psi}_{j-}^{\dag}({t^{\prime}})\breve{\psi}_{i-}(t)\rangle
&=& \langle(\breve{\psi}
_{jL}^{\dag}({t^{\prime}})-\breve{\psi}_{jR}^{\dag}({t^{\prime}}))(\breve{\psi}_{iL}(t)-
\breve{\psi}_{iR}(t))\rangle  \nonumber  \label{eq-5} \\
&=&\langle\breve{\psi}_{jL}^{\dag}({t^{\prime}})\breve{\psi}_{iL}(t)\rangle+\langle\breve{\psi}
_{jR}^{\dag}({t^{\prime}})\breve{\psi}_{iR}(t)\rangle  \nonumber \\
&-&\langle\breve{\psi}_{jR}^{\dag}({t^{\prime}})\breve{\psi}_{iL}(t)\rangle\!-\!\langle
\breve{\psi}_{jL}^{\dag}({t^{\prime}})\breve{\psi}_{iR}(t)\rangle  \nonumber \\
&\equiv& \langle\hat{\psi}_{j}^{\dag}({t^{\prime}})\hat{\psi}_{i}(t)\rangle+\langle\hat{\psi}
_{i}(t)\hat{\psi}_{j}^{\dag}({t^{\prime}})\rangle  \nonumber \\
&-&\langle\hat{\psi}_{j}^{\dag}({t^{\prime}})\hat{\psi}_{i}(t)\rangle\!-\!\langle
\hat{\psi}_{i}(t)\hat{\psi}_{j}^{\dag}({t^{\prime}})\rangle=0
\end{eqnarray}
Substituting from Eqs. (\ref{eq-4}) and (\ref{eq-5}) in
(\ref{eq-200}), we get Eq.(\ref{g--1}) for the retarded Green's
function.

Similarly,
\begin{eqnarray}
G^{ij}_{++}(t,{t^{\prime}})
&=& -i\langle\breve{T}\breve{\psi}
_{i+}(t)\breve{\psi}_{j+}^{\dag}({t^{\prime}})\rangle  \nonumber  \label{eq-20} \\
&=& -i\left[
\theta(t-{t^{\prime}})\langle\breve{\psi}_{i+}(t)\breve{\psi}
_{j+}^{\dag}({t^{\prime}})\rangle
 \theta({t^{\prime}}-t)\langle\breve{\psi}_{j+}^{\dag}({t^{\prime}}
)\breve{\psi}_{i+}(t)\rangle\right] .
\end{eqnarray}
Since
\begin{eqnarray}
\langle\breve{\psi}_{i+}(t)\breve{\psi}_{j+}^{\dag}\rangle
&=&\frac{1}{2}\langle(\breve{\psi}
_{iL}(t)+\breve{\psi}_{iR}(t))(\breve{\psi}_{jL}^{\dag}({t^{\prime}})+\breve{\psi}_{jR}^{\dag }
({t^{\prime}}))\rangle  \nonumber \\
&=&\frac{1}{2}\left[ \langle\breve{\psi}_{iL}(t)\breve{\psi}_{jL}^{\dag}({t^{\prime}}
)\rangle+\langle\breve{\psi}_{iL}(t)\breve{\psi}_{jR}^{\dag}({t^{\prime}})\rangle\right.
\nonumber \\
&+&\left. \langle\breve{\psi}_{iR}(t)\breve{\psi}_{jL}^{\dag}({t^{\prime}})\rangle
+\langle\breve{\psi}_{iR}(t)\breve{\psi}_{jR}^{\dag}({t^{\prime}})\rangle\right]
 =0
\end{eqnarray}
and
\begin{eqnarray}
\langle\breve{\psi}_{j+}^{\dag}({t^{\prime}})\breve{\psi}_{i+}(t)\rangle & =& \frac{1}{2}
\langle(\breve{\psi}_{jL}^{\dag}({t^{\prime}})+\breve{\psi}_{jR}^{\dag}({t^{\prime}}
))(\breve{\psi}_{iL}(t)+\breve{\psi}_{iR}(t))\rangle  \nonumber \\
& =&\frac{1}{2}\left[ \langle\breve{\psi}_{jL}^{\dag}({t^{\prime}})\breve{\psi}_{iL}(t)
\rangle+\langle\breve{\psi}_{jR}^{\dag}({t^{\prime}})\breve{\psi}_{iL}(t)\rangle\right.
\nonumber \\
& +& \left. \langle\breve{\psi}_{jL}^{\dag}({t^{\prime}})\breve{\psi}_{iR}(t)\rangle
+\langle\breve{\psi}_{jR}^{\dag}({t^{\prime}})\breve{\psi}_{iR}(t)\rangle\right]  \nonumber \\
&\equiv & \langle\{\hat{\psi}_{i}(t),\hat{\psi}_{j}^{\dag}({t^{\prime}})\}\rangle,
\end{eqnarray}
substituting these in Eq. (\ref{eq-20}), we get  Eq.(\ref{g++1}) for the
advanced Greens function.

We next turn to $G_{-+}$
\begin{eqnarray}
G^{ij}_{-+}(t,{t^{\prime}})
&=& -i\langle\breve{T}\breve{\psi}
_{i-}(t)\breve{\psi}_{j+}^{\dag}\rangle  \nonumber  \label{eq-7} \\
& =& -i\left[
\theta(t-{t^{\prime}})\langle\breve{\psi}_{i-}(t)\breve{\psi}
_{j+}^{\dag}({t^{\prime}})\rangle
-\theta({t^{\prime}}-t)\langle\breve{\psi}_{j+}^{\dag}({t^{\prime}}%
)\breve{\psi}_{i-}(t)\rangle\right].
\end{eqnarray}
Substituting for $\breve{\psi}_{-},\breve{\psi}_{+}$ in terms of $\breve{\psi}_{L}$ and
$\breve{\psi}_{R} $ gives
\begin{eqnarray}
\label{eq-8}
 \langle\breve{\psi}_{i-}(t)\breve{\psi}_{j+}^{\dag}({t^{\prime}})\rangle
 &=&\frac{1}{2}\langle(\breve{\psi}_{iL}(t)-\breve{\psi}_{iR}(t))(\breve{\psi}_{jL}^{\dag }({%
t^{\prime}})+\breve{\psi}_{jR}^{\dag}({t^{\prime}}))\rangle  \nonumber \\
& =& \frac{1}{2}\left[ \langle\breve{\psi}_{iL}(t)\breve{\psi}_{jL}^{\dag}({t^{\prime}}%
)\rangle+\langle\breve{\psi}_{iL}(t)\breve{\psi}_{jR}^{\dag}({t^{\prime}})\rangle\right.
\nonumber \\
&-&\left. \langle\breve{\psi}_{iR}(t)\breve{\psi}_{jL}^{\dag}({t^{\prime}})\rangle
-\langle\breve{\psi}_{iR}(t)\breve{\psi}_{jR}^{\dag}({t^{\prime}})\rangle\right]  \nonumber \\
&\equiv&\frac{1}{2}\left[ \langle\hat{\psi}_{i}(t)\hat{\psi}_{j}^{\dag}({t^{\prime}}%
)\rangle-\langle\hat{\psi}_{j}^{\dag}({t^{\prime}})\hat{\psi}_{i}(t)\rangle\right.
\nonumber \\
& +&\left. \langle\hat{\psi}_{i}(t)\hat{\psi}_{j}^{\dag}({t^{\prime}})\rangle
-\langle\hat{\psi}_{j}^{\dag}({t^{\prime}})\hat{\psi}_{i}(t)\rangle\right]
=\langle\lbrack\hat{\psi}_{i}(t),\hat{\psi}_{j}^{\dag}({t^{\prime}})]\rangle
\end{eqnarray}
Similarly we can simplify the other correlation function
\begin{eqnarray}
\label{eq-9}
\langle\breve{\psi}^\dag_{j+}({t^\prime})\breve{\psi}_{i-}(t)\rangle&=&
\frac{1}{2}\langle(\breve{\psi}^\dag_{jL}(\tp)+\breve{\psi}^\dag_{jR}(\tp))
(\breve{\psi}_{iL}(t)-\breve{\psi}_{iR}(t))\rangle\nonumber\\
&=& \frac{1}{2}\left[\langle \breve{\psi}^\dag_{jL}(\tp)\breve{\psi}_{iL}(t)\rangle
-\langle \breve{\psi}^\dag_{jR}(\tp)\breve{\psi}_{iR}(t)\rangle\right.\nonumber\\
&+&\left. \langle \breve{\psi}^\dag_{jR}(\tp)\breve{\psi}_{iL}(t)\rangle
-\langle\breve{\psi}^\dag_{jL}(\tp)\breve{\psi}_{iR}(t) \rangle
\right]\nonumber\\
&\equiv& \frac{1}{2}\left[\langle \hat{\psi}^\dag_{j}(\tp)\hat{\psi}_{i}(t)\rangle
-\langle\hat{\psi}_{i}(\tp)\hat{\psi}_{j}^\dag(t) \rangle\right.\nonumber\\
&+& \left.\langle\hat{\psi}_{j}^\dag(t)\hat{\psi}_{i}(\tp) \rangle
-\langle\hat{\psi}_{i}(\tp)\hat{\psi}_j^\dag(\tp) \rangle \right]\nonumber\\
&=&-\langle[\hat{\psi}_i(t),\hat{\psi}^\dag_j({t^\prime})]\rangle.
\end{eqnarray}
Equation (\ref{g-+1}) is obtained by substituting Eqs. (\ref{eq-9}) and (\ref{eq-8}) into (\ref{eq-7}).
This function is non-causal.

Finally we consider
\begin{eqnarray}
G^{ij}_{+-}(t) & =& -i\langle\breve{T}\breve{\psi}_{i+}(t)\breve{\psi}
_{j-}^{\dag}({t^{\prime}})\rangle  \nonumber  \label{eq-g+-} \\
& =&-i\left[
\theta(t-{t^{\prime}})\langle\breve{\psi}_{i+}(t)\breve{\psi}
_{j-}^{\dag}({t^{\prime}})\rangle
 - \theta({t^{\prime}}-t)\langle\breve{\psi}_{j-}^{\dag}({t^{\prime}}%
)\breve{\psi}_{i+}(t)\rangle\right] .
\end{eqnarray}
Since
\begin{eqnarray}
\langle\breve{\psi}_{i+}(t)\breve{\psi}_{j-}^{\dag}({t^{\prime}})\rangle
& =& \frac{1}{2}\langle(\breve{\psi}_{iL}(t)+\breve{\psi}_{iR}(t))(\breve{\psi}_{iL}^{\dag}({t^{\prime}%
})-\breve{\psi}_{iR}^{\dag}({t^{\prime}}))\rangle  \nonumber \\
& =& \frac{1}{2}\left[ \langle\breve{\psi}_{iL}(t)\breve{\psi}_{jL}^{\dag}({t^{\prime}}
)\rangle-\langle\breve{\psi}_{iL}(t)\breve{\psi}_{jR}^{\dag}({t^{\prime}})\rangle\right.\nonumber\\
&+&\left. \langle\breve{\psi}_{iR}(t)\breve{\psi}_{jL}^{\dag}({t^{\prime}})\rangle
-\langle\breve{\psi}_{iR}(t)\breve{\psi}_{jR}^{\dag}({t^{\prime}})\rangle\right]  \nonumber \\
& \equiv& \frac{1}{2}\left[ \langle\hat{\psi}_{i}(t)\hat{\psi}_{j}^{\dag}({t^{\prime}}
)\rangle+\langle\hat{\psi}_{j}^{\dag}({t^{\prime}})\hat{\psi}_{i}(t)\rangle\right.\nonumber\\
&-&\left. \langle\hat{\psi}_{i}(t)\hat{\psi}_{j}^{\dag}({t^{\prime}})\rangle-\langle
\hat{\psi}_{j}^{\dag}({t^{\prime}})\hat{\psi}_{i}(t)\rangle\right]=0.
\end{eqnarray}
Similarly it can be shown that $\langle\breve{\psi}_{j-}^{\dag}(t)\breve{\psi}_{i+}({t^{\prime}
})\rangle=0$. Equation (\ref{g+-1}) follows from Eq. (\ref{eq-g+-}).


\section{ Currents through
Molecular junctions with free-electron leads}\label{curr-exact}

In Sec. \ref{observables-current} we assumed a small quantum system $A$
coupled to an infinite system $B$. This corresponds to molecular wire or STM
experiments where a single molecule (or a quantum dot), the "system", is
coupled to two metal leads which are large compared to the
system. In this appendix we show how by adopting a free electron model of the leads,
the current can be calculated nonperturbatively
in the coupling with the leads.
The total Hamiltonian is
\begin{equation}
H_T=H_{S}+H_A+H_{B}+V.
\end{equation}
$H_A$ and $H_{B}$ are the free leads Hamiltonians given
in Eq. (\ref{eq-111}).
$V$ is the coupling of the system with the leads, Eq. (\ref{eq-112}).
The system Hamiltonian in this case is quite general and
can include electron-electron and/or electron-phonon interactions.
\begin{eqnarray}
\label{hamil-parts}
H_{S} &=&\sum_{i}\epsilon _{i}\hat{\psi} _{i}^{\dag }\hat{\psi} _{i}+
\sum_{ijkl}V_{ijkl}\hat{\psi}_{i}^{\dag }\hat{\psi}_{j}^{\dag }\hat{\psi}_{k}\psi
_{l} +\sum_{m}\Omega_{m}\hat{\phi}_{m}^{\dag }\hat{\phi}_{m}
+\sum_{ijm}\lambda _{ij}^{m}\hat{\psi}_{i}^{\dag }\hat{\psi}_{j}(\hat{\phi}
_{m}^{\dag }+\hat{\phi}_{m})\nonumber\\
\end{eqnarray}
where the fermion operators $\hat{\psi}^{\dag}(\hat{\psi} )$
satisfy the anti-commutation relations, Eq. \ref{eq-1a}.
Indices $(i,j,k,l)$ and $(m,n)$ represent the system electronic
states and system phonon states respectively.  $\epsilon _{i}$ and $\Omega _{m}$
are corresponding energies. $\lambda _{ij}$ and
$V_{ijkl}$ are the $e-p$ and $e-e$ interaction matrix
elements.
$\hat{\phi}_{m}^{\dag}(\hat{\phi}_{m})$ are phonon creation (annihilation) operator for
the $m$-th normal mode, with frequency $\Omega_{m}$. These operators
$\hat{\phi}$ satisfy the boson commutation relations, Eqs. (\ref{boson-commutation}).

The current from lead $A$ (left) to the system is defined as the rate of
change of the total charge of the lead,
\begin{eqnarray}  \label{eq-il}
I(t)\equiv\sum_l\frac{d}{dt} \langle\hat{N}_A\rangle=i\sum_l\langle[H_T,
\hat{N}_A(t)]\rangle
\end{eqnarray}
where $\hat{N}_{A}=e\sum_a\psi^{\dag}_{a}\psi_{a}$ is the number operator
for the left lead. We substitute the Hamiltonian and calculate the
commutator. $\hat{N}_{A}$ commutes with all terms in the Hamiltonian,
except the interaction $V$, Eq. (\ref{eq-112}). We obtain
\begin{eqnarray}  \label{eq-2}
I(t)= -2e\mbox{Im}\left\{ \sum_{ia}J_{ia}\langle\psi^{\dag
}_{i}(t)\psi_{a}(t)\rangle_{T}\right\}
\end{eqnarray}
where $\langle\cdots\rangle_{T}$ is the trace with respect to the total
(system+leads) density matrix. A Similar expression can be written for the
current from the right lead $B$ to the system by changing $a$ with
$b$ in Eq. (\ref{eq-2}). At steady state the two currents are same.

The Liouville space Green's functions are connected with their Hilbert space
counterparts in Appendix \ref{H-L-GF}. The current, Eq. (\ref{eq-2}), can be expressed
in terms of the Liouville space Greens function as
\begin{eqnarray}  \label{current-green}
I(t)=2e\mbox{Re}\{J_{ia}(t)G_{LR}^{ai}(t,t)\}.
\end{eqnarray}
In Eq.  (\ref{current-green}), the Green's function $G_{LR}^{ai}$ is defined in the
joint leads+system space. We compute $G_{LR}(t,{t^{\prime}})$ and then set $t={t^{\prime}}$.
In a recent work Rabani \cite{rabani} has used a numerical path-integral approach to
directly compute the Green's function $G_{LR}$ (hence the current).

We shall compute this Green's function using the equation of motion
technique. To that end, we construct the equation of motion for $\breve{\psi}_{a
\alpha}(t)(\alpha=L,R)$ using the Heisenberg equation for superoperators
\begin{eqnarray}
\label{eq-psil}
\frac{\partial}{\partial t}\breve{\psi}_{a\alpha}(t)=i\sqrt{2}[\breve{H}_-, \breve{\psi}_{a\alpha}(t)]
\end{eqnarray}
where on the right-hand side
\begin{eqnarray}
\label{tetradic}
[\breve{H}_-, \breve{\psi}_{a\alpha}(t)] = \breve{H}_-\breve{\psi}_{a\alpha}(t)-
\breve{\psi}_{a\alpha}(t)\breve{H}_-
\end{eqnarray}
and $\breve{H}_{-}=(\breve{H}_{L}-\breve{H}_{R})/\sqrt{2}$ can be expressed in terms of $\breve{\psi}_{L}$ and $
\breve{\psi}_{R}$ using Eq. (\ref{identity}).
Equation (\ref{eq-psil}) then gives
\begin{eqnarray}  \label{eq-psil-1}
i\frac{\partial}{\partial t}\breve{\psi}_{a\alpha}(t)=\epsilon_a\breve{\psi}_{a\alpha}(t)
+\sum_{i}J_{ai}^*(t)\breve{\psi}_{i\alpha}(t).
\end{eqnarray}

Taking the time derivative of Eq. (\ref{liouville-green}) with $x=l, y=i$
and $\mu=\alpha, \nu=\beta$, we get
\begin{eqnarray}  \label{new-eq-1}
i\frac{\partial}{\partial t}G_{\alpha\beta}^{ai}(t,{t^\prime})= \delta(t-{%
t^\prime})\delta_{\alpha\beta}\delta_{ai} +\langle \mathcal{T}\left(\frac{%
\partial}{\partial t}\hat{\psi}_{a\alpha}(t)\right)\hat{\psi}^\dag_{i\beta}({t^\prime}%
)\rangle.
\end{eqnarray}
The $\delta(t-{t^{\prime}})$ factor comes from the time derivation of the step
function (which originates from  $\breve{T}$) and the Kronecker delta functions
result from the commutation relations, Eqs. (\ref{eq-supercommu}). Using Eq. (\ref%
{eq-psil-1}) in (\ref{new-eq-1}) we get
\begin{eqnarray}  \label{eq-g}
(i\frac{\partial}{\partial t}-\epsilon_{a})G_{\alpha\beta}^{ai}(t,{t^\prime}%
)= \sum_{j}J_{aj}^*(t)G_{\alpha\beta}^{ji}(t,{t^\prime})
\end{eqnarray}
Note that $a\neq i$, since they belong to different regions of the total system. Thus
the first term on the r.h.s. in Eq. (\ref{new-eq-1}) does not contribute in this case. To
zeroth order in lead-system interaction, the Green's function for the leads
is defined as,
\begin{eqnarray}  \label{lead-green-def}
(i\frac{\partial}{\partial t}-\epsilon_{a}) g_{\alpha\beta}^{aa^\prime}(t,{%
t^\prime})=\delta(t-{t^\prime})\delta_{\alpha\beta}\delta_{aa^\prime}.
\end{eqnarray}
Substituting this in Eq. (\ref{eq-g}), we can write
\begin{eqnarray}  \label{matrix-eq}
G_{\alpha\beta}^{a i}(t,{t^\prime})=g_{\alpha{\beta^{\prime}}%
}^{aa^\prime}(t,t_1)J^*_{a^\prime {i^{\prime}}}(t_1) G^{i{i^{\prime}}}_{{%
\beta^{\prime}}\beta}(t_1,{t^\prime}).
\end{eqnarray}
For $G_{LR}$ this gives,
\begin{eqnarray}  \label{eq-glr}
G_{LR}^{ai}(t,{t^\prime})&=& J^*_{a^\prime{i^{\prime}}}(t_1)
\left(g_{LL}^{aa^\prime}(t,t_1)G_{LR}^{{i^{\prime}} i}(t_1,{t^\prime})
+g_{LR}^{aa^\prime}(t,t_1)G_{RR}^{{i^{\prime}} i}(t_1,{t^\prime}%
)\right).
\end{eqnarray}
The current, Eq. (\ref{current-green}), then becomes
\begin{eqnarray}  \label{curent-green-2}
I(t)\!\!&=&\!\! 2e\mbox{Re}\!\!\!\sum_{i{i^{\prime}} aa^\prime{%
\beta^{\prime}}}\!\!\int dt_1 J_{ia}J_{ja^\prime}^* g_{L{%
\beta^{\prime}}}^{aa^\prime}(t,t_1)G_{{\beta^{\prime}} R}^{ji}(t_1,t).
\end{eqnarray}
We define the Fourier transform
\begin{eqnarray}  \label{ft}
G(E_1,E_2)=\int dt_1 dt_2 \mbox{e}^{-i(E_1t_1+E_2t_2)}G(t_1,t_2).
\end{eqnarray}
Equation (\ref{curent-green-2}) then gives
\begin{eqnarray}  \label{eq-curr-11}
I(t)&=&2e\mbox{Re}\sum_{ijaa^\prime{\beta^{\prime}}}\int d\tilde{E}
J_{ai} J^*_{a^\prime {i^{\prime}}}\mbox{e}^{i(E_1+E_3)t}
g_{L{\beta^{\prime}}}^{ll^\prime}(E_1,E_2)G_{{\beta^{\prime}}
R}^{ji}(-E_2,E_3)
\end{eqnarray}
where $d\tilde{E}= dE_{1} dE_{2} dE_{3}/(2\pi)^{5}$. Using the
relations \cite{up-shaul-JCP}
\begin{eqnarray}
\label{green-relation}
G_{--} &=& G_{LL}- G_{LR},~ G_{++} = G_{LL} + G_{RL},~
G_{LL} + G_{RL}= G_{RR}+G_{LR}
\end{eqnarray}
equation (\ref{eq-curr-11}) can be transformed to
\begin{eqnarray}  \label{eq-curr-1}
I(t)&=&2e\mbox{Re}\sum_{ijaa^\prime}\int d\tilde{E} J_{ai}
J^*_{a^\prime {i^{\prime}}} \mbox{e}^{i(E_1+E_3)t}\nonumber\\
&\times&\left[g_{aa^\prime}^{--}(E_1,E_2)G_{LR}^{ji}(-E_2,E_3)
+ g^{a^\prime}_{LR}(E_1,E_2)G^{ij}_{++}(-E_2,E_3)\right].
\end{eqnarray}

We next compute the steady-state current ($t\to\infty$).
Integrating both sides over time, we obtain the (time-independent) current
\begin{eqnarray}
\label{eq-curr-3}
I&=&2e\mbox{Re}\sum_{ijaa^\prime} J_{ai} J^*_{a^\prime {i^{\prime}}%
}\int \frac{dE_1dE_2dE_3}{(2\pi)^3}
\left[g^{aa^\prime}_{--}(E_1,E_2)G_{LR}^{ji}(-E_2,-E_1)\right.
\nonumber \\
&&\quad\quad\quad+\left.g^{aa^\prime}_{LR}(E_1,E_2)G^{ij}_{++}(-E_2,-E_1)\right].
\end{eqnarray}
At steady state, since the one-particle Green's functions only depend on the
difference of their arguments, we can make the change of variables, $%
E_{1}-E_{2}=E$. Equation (\ref{eq-curr-3}) then gives
\begin{eqnarray}
\label{eq-il-a}
I&=&2e\mbox{Re}\sum_{i{i^{\prime}} aa^\prime}J_{ai}J^*_{a^\prime
{i^{\prime}}}\int \frac{dE}{2\pi} \left(g_{--}^{aa^\prime}(E)G_{LR}^{{
i^{\prime}} i}(E) +g_{LR}^{aa^\prime}(E)G_{++}^{{i^{\prime}} i}(E)\right).
\end{eqnarray}
Using the relations, ${G^{ij}_{--}}^{\ast}=G^{ji}_{++}$ and ${G_{LR}^{ij}}%
^{\ast}=-G_{LR}^{ji}$, we can write Eq. (\ref{eq-il-a}) as
\begin{eqnarray}  \label{eq-il-b}
I&=&e\sum_{i{i^{\prime}} aa^\prime}J_{ia}J_{a^\prime{i^{\prime}}}^*\int
\frac{dE}{2\pi} \left[G_{LR}^{i{i^{\prime}}}(E)\{2i\mbox{Im}%
g^{aa^\prime}_{--}(E)\}
- g_{LR}^{ll^\prime}(E) \{2i\mbox{Im}G_{i{i^{\prime}}}^{--}(E)\}
\right].\nonumber\\
\end{eqnarray}

From Eqs. (\ref{green-relation}), we have $-2i\mbox{Im}\{G_{--}\}
=G_{LR}+G_{RL}$. Substituting this in Eq. (\ref{eq-il-b}), we get
\begin{eqnarray}
\label{eq-final-1}
I&=&e\sum_{i{i^{\prime}} aa^\prime}J_{ia}J_{a^\prime{i^{\prime}}}^*\int
\frac{dE}{2\pi} \left[g_{LR}^{aa^\prime}(E)G_{RL}^{i{i^{\prime}}}(E)
-g_{RL}^{aa^\prime}(E)G_{LR}^{i{i^{\prime}}}(E) \right].
\end{eqnarray}
Since the leads are assumed to be in equilibrium, we can express their
Green's functions using the fluctuation-dissipation relations
\begin{eqnarray}  \label{fd}
g_{LR}^{aa^\prime}(E)&=&2\pi i f_A(E) \sigma_{aa}(E)\delta_{aa^\prime}
\nonumber \\
g_{RL}^{aa^\prime}(E)&=& 2\pi i (1-f_A(E)) \sigma_{aa}(E)\delta_{aa^\prime}
\end{eqnarray}
where $f_{A}(E)=[1+\mbox{exp}\{\beta(E-\mu_{A})\}]^{-1}$ and $\sigma
_{aa^{\prime}}(E)=-(1/\pi)\mbox{Im}\{G_{--}(E)\}$ are the Fermi function and the density of states (DOS)
for the left lead with chemical potential $\mu_{A}$. The current then
becomes
\begin{eqnarray}  \label{eq-final-2}
I=ie\sum_{i{i^{\prime}}}\Gamma_{i{i^{\prime}}}^A\int
\frac{dE}{2\pi} \left[f_A(E)G_{RL}^{i{i^{\prime}}}(E)-
(1-f_A(E))G_{LR}^{i{i^{\prime}}}(E) \right]
\end{eqnarray}
where $\Gamma_{i{i^{\prime}}}^{A}=\sum_{aa^{\prime}}2\pi J_{ia}J_{{i^{\prime}%
} a^{\prime}}^{*}\sigma_{aa^{\prime}}$ and the DOS of the leads is assumed
to be energy independent in the relevant energy range (chemical potential difference
of the two leads). This can also be expressed in terms of the self-energies due to
the interaction with the left lead $\Sigma_{\alpha\beta}$
\begin{eqnarray}
\label{eq-final-2-new}
I=e\sum_{i{i^{\prime}}}\int \frac{dE}{2\pi} \left[{
\Sigma^{i{i^{\prime}}}_{LR}}(E)G_{RL}^{i{i^{\prime}}}(E)-{\Sigma^{i{
i^{\prime}}}_{RL}}(E)G_{LR}^{i{i^{\prime}}}(E) \right]
\end{eqnarray}
where $\Sigma^{i{i^{\prime}}}_{LR}(E)=\Gamma_{ii^\prime}^Af_A(E)$ and $%
\Sigma^{i{i^{\prime}}}_{RL}(E)=\Gamma_{ii^\prime}^A[1-f_A(E)]$.
Equation (\ref{eq-final-2-new}) is same as Eq. (\ref{nn-33}).

A similar expression can be obtained for the current from lead $B$ to the
system by replacing $a$ with $b$ in Eq. (\ref{eq-final-2}).
\begin{eqnarray}
\label{eq-final-3}
I^\prime=ie\sum_{i{i^{\prime}}}\Gamma_{i{i^{\prime}}
}^B\int \frac{dE}{2\pi} \left[f_B(E)G_{RL}^{i{i^{\prime}}%
}(E)-(1-f_B(E))G_{LR}^{i{i^{\prime}}}(E) \right].
\end{eqnarray}
At steady state $I=-I^\prime$ and the steady state current $I_{s}$ can be
written as, $I_{s}=xI-(1-x)I^\prime$ for any value of $x$. Using Eqs. (\ref%
{eq-final-2}) and (\ref{eq-final-3}), we obtain
\begin{eqnarray}  \label{eq-is-4}
I_s&=&ie\sum_{i{i^{\prime}}}\int \frac{dE}{2\pi} \left\{[xf_A(E)\Gamma^A_{i{
i^{\prime}}}-(1-x)f_B(E)\Gamma^B_{i{i^{\prime}}}]\right.  \nonumber \\
&\times&\left.(G_{RL}^{i{i^{\prime}}}(E)+G_{LR}^{i{i^{\prime}}}(E))
-[x\Gamma^A_{i{i^{\prime}}} -(1-x)\Gamma^B_{i{i^{\prime}}}]G_{LR}^{i{
i^{\prime}}}(E) \right\}  \nonumber \\
\end{eqnarray}
This can be simplified further by using, $-2i\mbox{Im}\{G_{--}%
\}=G_{LR}+G_{RL}$, to get
\begin{eqnarray}
\label{eq-is-5}
I_s&=&-ie\sum_{i{i^{\prime}}}\int \frac{dE}{2\pi} \left\{2i[xf_A(E)
\Gamma^A_{i{i^{\prime}}}-(1-x)f_B(E)\Gamma^B_{i{i^{\prime}}}]\right.  \nonumber
\\
&\times&\left.\mbox{Im}G^{--}_{i{i^{\prime}}}(E) +[x\Gamma^A_{i{i^{\prime}}%
}-(1-x)\Gamma^B_{i{i^{\prime}}}]G_{LR}^{i{i^{\prime}}}(E) \right\}.
\end{eqnarray}

This is the general exact formal expression for the current that includes
both the left and right lead properties. As a check let us assume no
external bias and set, $f_{A}=f_{B}=f$. Equation (\ref{eq-is-5}) then reduces to
\begin{eqnarray}
\label{eq-is-6}
I_s&=&-ie\sum_{i{i^{\prime}}}\int \frac{dE}{2\pi} [x\Gamma^A_{i{i^{\prime}}
}-(1-x)\Gamma^B_{i{i^{\prime}}}]
[2if(E)\mbox{Im}G_{--}^{i{i^{\prime}}}(E) +G_{LR}^{i{i^{\prime}}}(E)]
\end{eqnarray}
At equilibrium, $G_{LR}(E)=-2if(E)\mbox{Im}G_{--}(E)$ (which is the FD
theorem)\cite{haug-jauho}, and $I_{s}$ vanishes, as it should.

Assuming that the coupling elements are real, $\Gamma_{ij}=\Gamma_{ji}$, Eq.
(\ref{eq-final-3}) can be expressed in a simple form in terms of
retarded and correlation Greens functions as
\begin{eqnarray}  \label{current-pp-mp}
I=ie\sum_{i{i^{\prime}}}\Gamma_{i{i^{\prime}}}^A\int \frac{dE%
}{2\pi} \left[i(1-2f_A(E))\mbox{Im}G_{--}^{i{i^{\prime}}}(E)+\frac{1}{2%
}G^{-+}_{i{i^{\prime}}}(E)\!\right].
\end{eqnarray}

When the left and the right lead-system couplings satisfy $\Gamma_{i{
i^{\prime}}}^{A}=\lambda\Gamma_{i{i^{\prime}}}^{B}$, the current $I_{s}$ can
be expressed solely in terms of the imaginary part of the retarded Greens
function. Choosing $x=1/(1+\lambda)$ in Eq. (\ref{eq-is-5}) gives
\begin{eqnarray}
\label{steady-current}
I_s=2e\sum_{i{i^{\prime}}}\frac{\Gamma^A_{i{i^{\prime}}}\Gamma^B_{i{%
i^{\prime}}}}{\Gamma^A_{i{i^{\prime}}}+\Gamma^B_{i{i^{\prime}}}} \mbox{Im}%
\int\frac{dE}{2\pi} (f_A(E)-f_B(E))G_{--}^{i{i^{\prime}}}(E).
\end{eqnarray}
This can be recast as,
\begin{eqnarray}
\label{new-lb-form}
I_s=2e
\int\frac{dE}{2\pi} (f_B(E)-f_A(E))\sum_{i{i^{\prime}}}|S_{i{i^{\prime}}}(E)|^2.
\end{eqnarray}
where
\begin{eqnarray}
\label{new-s}
|S_{ii^\prime}(E)|^2 =
\frac{\Gamma^A_{i{i^{\prime}}}\Gamma^B_{i{i^{\prime}}}}{\Gamma^A_{i{i^{\prime}}}+\Gamma^B_{i{i^{\prime}}}}
\mbox{Im}G^{ii^\prime}_{--}(E)
\end{eqnarray}
Equation (\ref{new-lb-form}) has the LB form given in  Eq. (\ref{nn-44}).
From Eq. (\ref{steady-current}) it is easy to see that the steady state
current vanishes when the external bias is turned off ($f_{A}=f_{B}$) or one
of the leads is disconnected (so that $\Gamma^{A}=0$ or $\Gamma^{B}=0$). We
note that the condition $\Gamma_{i{i^{\prime}}}^{A}=\lambda\Gamma _{i{%
i^{\prime}}}^{B}$ trivially holds when the leads are coupled to a single system
orbital.

When the system is modeled as non-interacting electron system and the
system-lead interactions are small, we can use
\begin{equation}
G^{i{i^{\prime }}}_{--}(E)=\delta _{i{i^{\prime }}}\frac{1}{E-\epsilon
_{i}+i\eta}
\end{equation}
where $\epsilon _{i}$ is the single electron energy. Substituting this in Eq.
(\ref{steady-current}) then gives
\begin{equation}
\label{nmn}
I_{s}=2e\sum_{i}\!\frac{\Gamma _{ii}^{A}\Gamma _{ii}^{B}}{\Gamma
_{ii}^{A}+\Gamma _{ii}^{B}}(f_{B}(\epsilon _{i})-f_{A}(\epsilon _{i})).
\end{equation}
This coincide with the quantum master equation
appraoch\cite{up-max-shaulPRB1}.

In general, the steady state current is given by Eq. (\ref{eq-final-2-new}) which
involves the Green's functions $G_{LR}$ and $G_{RL}$. These can be calculated by solving the
matrix Dyson equation self-consistently. When the coupling to the left and the right
leads are proportional, $\Gamma^A=\lambda\Gamma^B$,  Eq. (\ref{eq-final-2-new}) can be
recast in the simpler form, Eq. (\ref{steady-current}), which involves properties of
both the leads. Finally, when the molecule is modeled as
noninteracting electron system, the current is simply the sum of independent contributions from
each orbital (Eq. (\ref{nmn})).


\section{Connecting Eq. (\ref{cum-gf}) with Eq. (\ref{gf-final})}\label{mod-evol}

The time evolution operator ${\cal G}^I$ in Eq. (\ref{cum-gf}) is
given by\cite{two-point}
\begin{equation}
\label{v-tau}
\mathcal{G}_{I}(\lambda,\Lambda,t)=\breve{T}\mbox{exp}\left\{
-i\sqrt{2}\int_{-\infty }^{t}d\tau \breve{V}^I_{-}(\gamma _{\alpha }(t);\tau
)\right\}
\end{equation}
where
\begin{eqnarray}
\label{v-tau-1}
\breve{V}_{-}(\gamma _{\alpha }(t);\tau )&=&\frac{1}{\sqrt{2}}[\breve{V}_{L}(\gamma
_{L}(\tau);\tau )-\breve{V}_{R}(\gamma _{R}(\tau );\tau )]
\end{eqnarray}
\begin{eqnarray}
\label{v-tau-2}
\breve{V}_{\alpha }(\gamma _{\alpha }(t);t)&=&e^{-i\gamma _{\alpha
}(t)}\breve{H}_{AB\alpha }^{\prime }(t)+h.c.
\end{eqnarray}

Note that for $\gamma _{\alpha }(t)=0$, $\mathcal{G}_{I}(\gamma
_{\alpha }(t);t,t_{0})$ is simply the time evolution operator for
the density matrix. The time dependence in Eq. (\ref{v-tau}) is due
to the interaction picture with respect to Hamiltonian
$\hat{H}_{0}=\hat{H}_{A}+\hat{H}_{B}$ and $\breve{H}_{AB\alpha
}^{\prime }=\sum_{ab}J_{ab}({\hat{\psi}^{\dagger
}}_{a}{\hat{\psi}} _{b})_{\alpha}$. The parameters $\gamma
_{\alpha }(t)\equiv\theta (t)\gamma _{\alpha }$ with
\begin{equation}
\gamma _{L}=\Gamma +\frac{\gamma }{2},~~~~\gamma _{R}=\Gamma
-\frac{\gamma }{ 2}.
\end{equation}
$\gamma _{\alpha }$ is an auxiliary field in Liouville space which
modifies the tunneling between the two sub-systems and acts
differently on the ket and the bra of the density matrix. A similar
approach was used \cite {up-shaul-JCP} to extend the well known GW
equations \cite{hedin,onida-RMP,gunnarsson-Review} to non-equilibrium systems. The GF,
$Z (\lambda ,\Lambda ,t)$ can be computed perturbatively in the
tunneling matrix elements, $J_{AB}$. Since $ \langle
\hat{V}_{-}(\tau )\rangle =0$ (particle conservation in individual
noninteracting systems),to lowest (second) order we obtain,
\begin{equation}
\label{cum-gf-11}
Z(\lambda ,\Lambda ,t)=-\int_{t_{0}}^{t}\int_{t_{0}}^{t} d\tau
_{1}d\tau _{2}\langle \breve{T}\breve{V}_{-}(\gamma _{\alpha }(\tau
_{1});\tau _{1})\breve{V}_{-}(\gamma _{\alpha }(\tau _{2});\tau
_{2})\rangle
\end{equation}

By substituting Eqs. (\ref{v-tau-1}) and (\ref{v-tau-2}) for $\breve{V}_{-}(\tau )$ and noting
that $ \langle \breve{H}_{AB\alpha }(\tau _{1})\breve{H}_{AB\beta
}(\tau _{2})\rangle $ vanishes (particle conservation; trace is over
the product states), we obtain Eq. (\ref{gf-final}).

\section{Self-energy for Electron-electron interactions}\label{se-ee}

The $e-e$ interactions in the system are represented by the Hamiltonian.
\begin{eqnarray}
\hat{H}_{int} = \sum_{ijkl}V_{ijkl}{\hat{\psi}^\dagger}_i{\hat{\psi}^\dagger}%
_j{\hat{\psi}}_k{\hat{\psi}}_l
\end{eqnarray}
with
\begin{eqnarray}  \label{vijkl}
V_{ijkl}=e\int\int dx d x^\prime
\varphi_i^*(x)\varphi_k(x)\frac{1}{|x-x^\prime|} \varphi_j^*(x^\prime)\varphi_l(x^\prime)
\end{eqnarray}
and $\varphi_{i}$ is the basis set wavefunction at site $i$. Note that $%
V_{ijkl}=V_{jilk}$. In this section all the superoperator indices
shall correspond to $+$ and $-$ operators.

The corresponding superoperator in the PTSA representation is given by
\begin{eqnarray}  \label{int-super}
L_{int}&=&\sum_{ijkl}\sum_{\alpha\beta\ap\bp}
V_{ijkl}\left[{{\breve{\psi}^\dagger}}_{i\alpha}{{\breve{\psi}^\dagger}}_{j\beta}
{{\breve{\psi}}}_{k{\ap}}\breve{\psi}_{l\bp}
+(-1)^{p+1}\breve{\psi}_{l\bp}\breve{\psi}_{k\ap}\breve{\psi}^\dagger_{j\beta}
\breve{\psi}^\dagger_{i\alpha} \right]
\end{eqnarray}
where $p$ is the number of "minus" operators in the second term
inside the bracket. Substituting this in Eq.
(\ref{interaction-self-energy}) and expanding the exponential, the
self-energy $\Sigma_{\alpha\beta}$ can be evaluated perturbatively in
$e-e$ interaction, $V_{ijkl}$. To first order we get,
\begin{eqnarray}
\label{self-energy-new}
\Sigma^{ij}_{\alpha\beta}(t,\tp)\!&=&\!\!
\frac{-i}{2}\delta(t-\tp)\!\!\sum_{\ap\bp}\sum_{i^\prime k^\prime}
W^{{i^{\prime}}i{k^{\prime}} j}_{\ap\alpha\bp\beta}
G^{{k^{\prime}}{i^{\prime}}}_{\bp\ap}(t,t)
\end{eqnarray}
where
\begin{eqnarray}  \label{new-w}
W^{{i^{\prime}} i{k^{\prime}} j}_{{\kappa^{\prime}}\kappa{\eta^{\prime}}%
\eta}= (V_{{i^{\prime}} i{k^{\prime}}
j}-V_{i{i^{\prime}}{k^{\prime}} j})[1+(-1)^{p+1}]
\end{eqnarray}
with $p$ the number of "minus" indices $\kp,\kappa,\eta^\prime\eta$ of $W$.
The electron-phonon self-energy can be calculated similarly\cite{up-shaul-prb1}.


\section{Self-energy for a molecular-wire connected to free electron leads}\label{se-ml}

Here we compute the Green's functions for a molecule attached to two leads. The
leads are assumed to be non-interacting electron system which remain at equilibrium
with their respective chemical potentials. We also ignore the many-body interactions
in the molecule. For this model, the interaction with the leads can be computed
exactly within the Green's function approach.
The total Hamiltonian is given by, Eq. (\ref{hamil-2}).

The EOM for the Green's function
$G^{ij}_{\alpha\beta}(t,\tp)$ is obtained as in Eq. (\ref{new-eq-1}).
\begin{eqnarray}
\label{seq-1}
i\frac{\partial}{\partial t}G_{\alpha\beta}^{ij}=\delta(t-\tp)\delta_{\alpha\beta}\delta_{ij}
+\left\langle \breve{T}\frac{\partial}{\partial t}\breve{\psi}_{i\alpha}(t)
\breve{\psi}^\dag_{j\beta}(\tp)\right\rangle.
\end{eqnarray}
Using the EOM for superoperator $\breve{\psi}_{i\alpha}$ in the Heisenberg picture
\begin{eqnarray}
\label{seq-2}
i\frac{\partial}{\partial t}\breve{\psi}_{i\alpha}(t) = \epsilon_i\breve{\psi}_{i\alpha}(t)+
\sum_{x\in A,B}J_{ix}\breve{\psi}_{x\alpha},
\end{eqnarray}
Eq. (\ref{seq-1}) can be expressed as (repeated indices are summed over),
\begin{eqnarray}
\label{seq-3}
\left(i\frac{\partial}{\partial t}-\epsilon_i\right)G_{\alpha\beta}^{ij}(t,\tp)
=\delta(t-\tp)\delta_{\alpha\beta}\delta_{ij}+\Sigma_{\alpha\bp}^{ik}(t\tp_1)G_{\bp\beta}^{kj}(\tp_1,\tp).
\end{eqnarray}
In Eq. (\ref{seq-3}), the self-energy ($\sigma$) due to molecule-lead couplings is defined as
\begin{eqnarray}
\label{seq-4}
\Sigma_{\alpha\bp}^{ik}(t\tp_1)G_{\bp\beta}^{kj}(\tp_1,\tp)=-i\sum_{x\in A,B}J_{ix}
\left\langle\breve{T}\breve{\psi}_{x\alpha}(t)\breve{\psi}_{j\beta}^\dag(\tp)\right\rangle.
\end{eqnarray}

Note that in Eq. (\ref{seq-4}) on the rhs the self-energy is given in terms of the cross-correlation
Green's function defined in the joint space of the lead and the molecule. We wish to express it
in using the Green's functions
of the individual systems non-perturbatively. For this purpose, starting from the EOM for the
lead superoperator $\breve{\psi}_{x\alpha}$, we construct the EOM for the joint Green's function.
\begin{eqnarray}
\label{seq-5}
\left(i\frac{\partial}{\partial t}-\epsilon_x\right)G_{\alpha\beta}^{xj}(t,\tp)
=\sum_{k}J_{xk}G_{\alpha\beta}^{kj}(t,\tp).
\end{eqnarray}
This can also be recast as
\begin{eqnarray}
\label{seq-6}
G_{\alpha\beta}^{xj}(t,\tp)
=\sum_{k\xp}g_{\alpha\bp}^{x\xp}(t,\tp_1)
J_{\xp k}G_{\bp\beta}^{kj}(\tp_1,\tp)
\end{eqnarray}
where $g_{\alpha\beta}^{x\xp}$ is the Green's function for the free (in absence of the molecule) leads
defined as,
\begin{eqnarray}
\label{seq-7}
\left(i\frac{\partial}{\partial t}-\epsilon_x\right)g_{\alpha\beta}^{x\xp}(t,\tp)
=\delta(t-\tp)\delta_{\alpha\beta}\delta_{x\xp}.
\end{eqnarray}
Substituting Eq. (\ref{seq-6}) in (\ref{seq-4}), and using the identity
$G_{\alpha\bp}^{ij}(t,\tp_1)G_{\bp\beta}^{jk}(\tp_1,\tp)=\delta_{\alpha\beta}\delta_{ik}$ ,
the self-energy can be written as
\begin{eqnarray}
\label{seq-8}
\Sigma_{\alpha\beta}^{ij}(t,\tp)=-i\sum_{x,\xp}J_{ix}g_{\alpha\beta}^{x\xp}(t,\tp)J_{\xp j}.
\end{eqnarray}
Since $g_{\alpha\beta}^{x\xp}$ is the Green's function for the independent leads, it is
diagonal in $x,\xp$, $g_{\alpha\beta}^{x\xp}=\delta_{x\xp}g_{\alpha\beta}$. Transforming to
frequency domain, Eq. (\ref{seq-8}) then gives,
\begin{eqnarray}
\label{seq-9}
\Sigma_{\alpha\beta}^{ij}(E)= -i\sum_{x\in A,B}J_{ix}g_{\alpha\beta}^{xx}(E)J_{x j}.
\end{eqnarray}

The self-energy can be finally obtained using the expressions for the lead Green's
functions $g_{LR}$ and $g_{RL}$ given in Eqs. (\ref{fd}). The Green's function
$g_{LL}=g_{--}+g_{LR}$ [Eq. (\ref{green-relation})], where $g_{--}$ can be obtained from Eq. (\ref{last}), and
$g_{RR}=g_{LL}+g_{RL}-g_{LR}$ [Eq. (\ref{green-relation})]. This defines all elements of the self-energy matrix
which appears in Eqs. (\ref{nn-33}) and (\ref{eq-final-2-new}).


\end{document}